\documentclass[11pt,a4paper,oneside]{article}

\usepackage{jheppub}
\usepackage{color}
\usepackage{graphicx}
\usepackage{wrapfig,enumerate,slashed}

\usepackage{footmisc}
\usepackage{amsmath}
\usepackage{wasysym} 
\usepackage{graphicx}
\usepackage{subcaption}
\usepackage{color}
\usepackage{comment}
\usepackage{appendix}
\usepackage{slashed}

\newcommand{\be}{\begin{eqnarray}}
\newcommand{\ee}{\end{eqnarray}}

\newcommand{\bee}{\begin{eqnarray}}
\newcommand{\eee}{\end{eqnarray}}
\newcommand{\beeq}{\begin{equation}}
\newcommand{\eeeq}{\end{equation}}

\makeatletter
\gdef\@fpheader{}
\makeatother
\begin{document}

\title{Towards a Quantum Computing Algorithm for Helicity Amplitudes and Parton Showers}

\author[a]{Khadeejah Bepari,}
\author[b]{Sarah Malik,}
\author[a]{Michael Spannowsky}
\author[b]{and Simon Williams}

\affiliation[a]{Institute for Particle Physics Phenomenology, Department of Physics, Durham University, Durham DH1 3LE, U.K.}
\affiliation[b]{High Energy Physics Group, Blackett Laboratory, Imperial College,
Prince Consort Road, London, SW7 2AZ, United Kingdom}

\emailAdd{khadeejah.bepari@durham.ac.uk}
\emailAdd{sarah.malik@imperial.ac.uk}
\emailAdd{michael.spannowsky@durham.ac.uk}
\emailAdd{s.williams19@imperial.ac.uk}

\abstract{The interpretation of measurements of high-energy particle collisions relies heavily on the performance of full event generators. By far the largest amount of time to predict the kinematics of multi-particle final states is dedicated to the calculation of the hard process and the subsequent parton shower step. With the continuous improvement of quantum devices, dedicated algorithms are needed to exploit the potential quantum computers can provide. We propose general and extendable algorithms for quantum gate computers to facilitate calculations of helicity amplitudes and the parton shower process. The helicity amplitude calculation exploits the equivalence between spinors and qubits and the unique features of a quantum computer to compute the helicities of each particle involved simultaneously, thus fully utilising the quantum nature of the computation. This advantage over classical computers is further exploited by the simultaneous computation of s and t-channel amplitudes for a 2$\rightarrow$2 process. The parton shower algorithm simulates collinear emission for a two-step, discrete parton shower. In contrast to classical implementations, the quantum algorithm constructs a wavefunction with a superposition of all shower histories for the whole parton shower process, thus removing the need to explicitly keep track of individual shower histories. 
Both algorithms utilise the quantum computer’s ability to remain in a quantum state throughout the computation and represent a first step towards a quantum computing algorithm to describe the full collision event at the LHC.}

\preprint{IPPP/20/41}

\maketitle


\section{Introduction}
\label{sec:intro}
Modern collider experiments such as the Large Hadron Collider (LHC) at CERN depend heavily on the modelling of particle collisions and simulations of detector response to examine physics processes within the experiments. This modelling is used to construct different possible outcomes from particle collisions, used both for the identification of certain physical processes, and for the construction of event backgrounds. Consequently, such simulations play a crucial role in modern high energy physics, and are usually carried out by Monte Carlo event generators such as \textsc{Pythia}~\cite{Pythia}, \textsc{Herwig}~\cite{Herwig} and \textsc{Sherpa}~\cite{Gleisberg:2003xi}. 

The theoretical description of LHC events can be highly complex. In a typical event, hundreds of particles are produced as a result of the evolution of an event from the collision of two protons to the formation of long-lived hadrons, leptons and photons. The collision process can be separated into several stages. The protons consist of many partons, each carrying a fraction of the total proton energy. When protons collide, two of their partons can interact with each other via a large momentum transfer, thereby giving rise to the so-called hard interaction. In this part of the collision, large interaction scales are probed, possibly accessing new physics. However, if color-charged particles are produced during the hard interaction process, they are likely to emit further partons. This results in a parton shower, providing a mechanism that evolves the process from the hard interaction scale down to the hadronisation scale $\mathcal{O}(\Lambda_{\mathrm{QCD}})$, where non-perturbative processes rearrange the partons into colour-neutral hadrons.

The hard interaction and the parton shower are the two parts of the event evolution that can be described perturbatively and largely independently of non-perturbative processes, as a result of the factorisation theorem \cite{Collins:1989gx}. In addition, they are by far the most time-consuming parts of the event simulation and pose, therefore, the bottleneck in the generation of pseudo-data for ongoing measurements at the LHC.

While a speed improvement in calculating the hard process and the parton shower is crucial for the interpretation of high-energy collision experiments, the conceptual methods used to calculate either of these two parts are distinctly different. For a mathematical description of the hard interaction, scattering matrix elements are calculated, which nowadays rely on helicity amplitude methods to cope with the ever-increasing complexity of the partonic scattering process \cite{Parke:1986gb,Berends:1987me}.
Instead, the parton shower is technically implemented through a Markov chain algorithm ordered in some measure of showering time $t$, where splitting functions define the probability for a parton to branch into two partons and Sudakov factors \cite{Sudakov:1954sw} determine the probability for the system not to change between two shower times\footnote{For more details see \cite{Buckley:2011ms} and references therein.} $t_\mathrm{in}$ and $t_\mathrm{end}$.
Recent developments in combining helicity amplitudes with the parton shower have shown to improve the theoretical description of scattering events including multiple jets \cite{Catani:2001cc,Mangano:2001xp,Lonnblad:2001iq,Hoche:2015sya, Fischer:2017yja,Fischer:2017htu}, in hypothesis testing \cite{Prestel:2019neg,Soper:2014rya} and in particular in the construction of spin-dependent parton showers \cite{Richardson:2018pvo}.

With practical quantum computers becoming available, there has been growing interest in harnessing the power and advantages that these machines may provide. This interest extends to applying the abilities of quantum computers to describe processes in field theories, with the hope of exploiting the intrinsic `quantumness' of these novel machines to calculate quantum phenomena efficiently. Current quantum computers are divided into two classes: quantum annealers and universal gate quantum computers (GQC). The former is based on the adiabatic theorem of quantum mechanics to find the ground state of a complex system. Quantum annealers perform continuous-time quantum computations and are therefore well-suited to study the dynamics of quantum systems, even quantum field theories \cite{Abel:2020ebj,Abel:2020qzm}, and in solving optimisation problems, e.g. applied to Higgs phenomenology \cite{MottQuantum}. However, they are not universal.
Despite their severe limitation due to the relatively small number of qubits of current machines, GQC are a popular choice for the implementation of algorithms to calculate multi-particle processes \cite{Jordan:2011ci,Garcia-Alvarez:2014uda,Jordan:2014tma,Jordan:2017lea,Preskill:2018fag,bauer2019quantum,Moosavian:2019rxg,Alexandru:2019ozf, Alexandru:2019nsa, Lamm:2019uyc, Lamm:2020jwv}, often with field theories mapped onto a discrete quantum walk  \cite{Marque-Martin:2018PRA,Arrighi:2018PRA,Jay:2019PRA,DiMolfetta:2020QIP} or a combined hybrid classical/quantum approach \cite{Lamm:2018siq,Harmalkar:2020mpd,Wei:2019rqy,Matchev:2020wwx}.

Here, we aim to provide a first step towards a generic implementation of quantum algorithms, applicable to QGC devices, for the most time-consuming parts of the event generation in high-energy collisions, i.e. the calculation of the hard process in terms of helicity amplitudes and the simulation of the parton shower\footnote{A first implementation of a parton shower algorithm was provided in \cite{bauer2019quantum}, where interference effects in the parton shower evolution were studied.}.

\begin{figure}[!h]
\centering
  \includegraphics[width=0.8\textwidth]{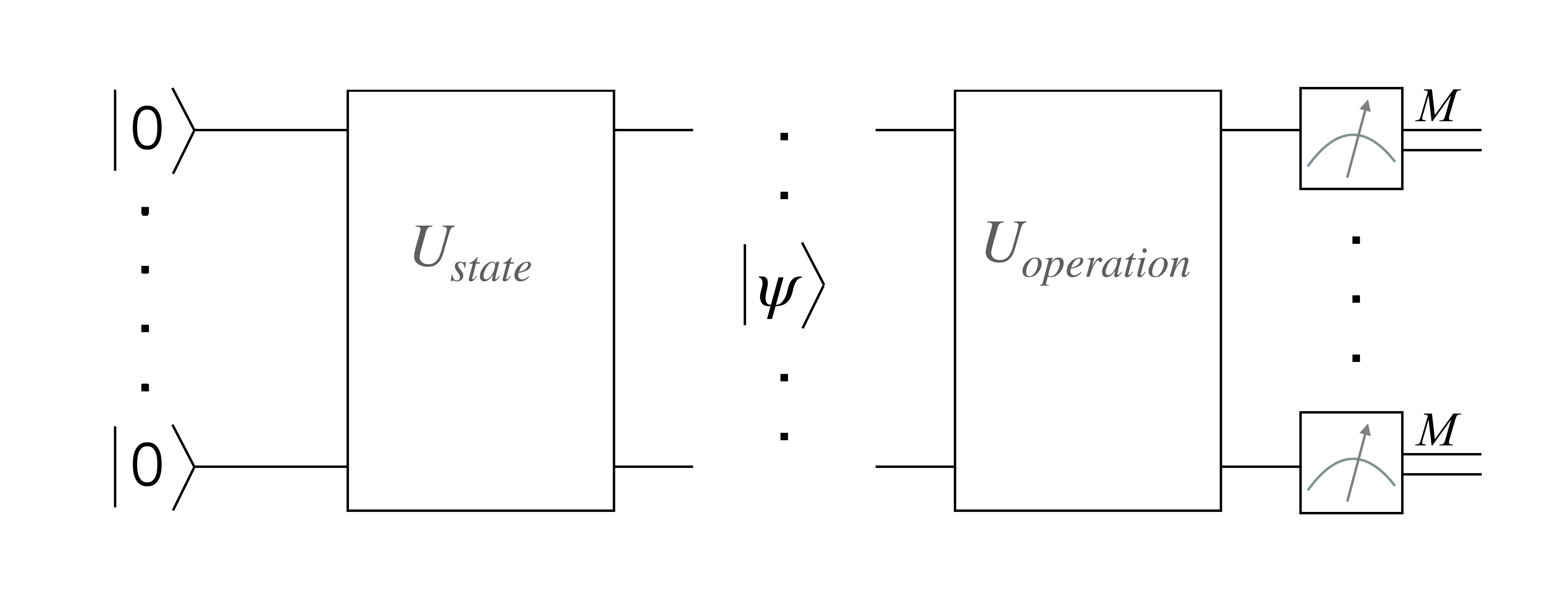}
\caption{Schematic setup of generic quantum computing calculations with the following steps: (i) encoding of the initial state, (ii) the application of (unitary) operations and (iii) the measurement of the transformed state.}
\label{fig:encoding}
\end{figure}

As depicted in Fig.~\ref{fig:encoding}, QC calculations proceed in general in three stages: i) encoding of the initial state, i.e. an initial wavefunction, using a specific representation of the problem, ii) applying unitary operations on this state, which on a GQC is realised through circuits, and iii) measuring a specific property of interest, i.e. a projection onto the final state vector. 

Following this structure, we will elucidate how the calculation of a hard process in terms of helicity amplitudes or the parton shower can be performed using a GQC. Specifically, we use the IBM Q Experience~\cite{IBMQ}, which provides access to a range of public access quantum computers and a 32-qubit Quantum Simulator~\cite{32_sim}. We have designed the circuits with a focus on limiting the number of qubits needed to perform the calculations. While our code can be run on a real quantum device, the current quantum machines cannot outperform classical computers. The quantum circuits presented here, therefore, serve as a template and nucleus for future developments.

This paper is organised as follows: in Sec.~\ref{sec:helicity}, we motivate and detail the implementation of our QC algorithm for helicity amplitudes for a 1$\rightarrow$2 process and a 2$\rightarrow$2 scattering process, Sec.~\ref{sec:parton} contains the description of our 2-step parton shower algorithm and Sec.~\ref{sec:conclusions} offers a summary and conclusions.

\section{Helicity amplitude algorithm}
\label{sec:helicity}

Scattering processes are calculated using conventional techniques by squaring the scattering amplitude and then performing a sum of all possible helicity processes using trace techniques. For a process with $N$ possible Feynman diagrams, this results in $N^2$ terms in the squared amplitude. Therefore, for processes with a large number of Feynman diagrams, such calculations become extremely complicated. In contrast, helicity amplitude calculations provide a more efficient way of calculating such processes, as one calculates the amplitude for a specific helicity setup. The different helicity combinations do not interfere, and therefore the full amplitude can be obtained by summing the squares of all possible helicity amplitudes.

Helicity amplitude calculations are based on the manipulation of helicity spinors. As the Lorentz group Lie algebra can be written as the direct sum of two $SU(2)$ sub-algebras, i.e. $so(3,1)= su(2) \oplus  su(2)$, there are two specific complex representations each specified by two degrees of freedom which solve the massless Weyl equation: a right-handed Weyl spinor, associated with the representation $(\frac{1}{2},0)$, and a left-handed Weyl spinor, associated with the representation $(0,\frac{1}{2})$. 
Consequently and for concreteness, the helicity spinor $\vert p \rangle^{\dot{a}}$ for a massless state can be chosen to be expressed as

\begin{align}
\vert p \rangle^{\dot{a}}=\sqrt{2E}\begin{pmatrix}
\cos\frac{\theta }{2}\\ 
\sin\frac{\theta }{2}e^{i\phi }
\end{pmatrix},
\label{eq:spinor}
\end{align}
associated with momentum $p^\mu$ and energy $E$, such that $p^\mu p_\mu=-m^2$ using the  $\eta _{\mu \nu }$=diag(-1, +1, +1, +1) metric convention.
This spinor is parametrised by the angles $\theta$ and $\phi$, where the other spinors $\langle p \vert_{\dot{a}}$, $| p ]_a$ and $[ p |^a$ are related by $p_{a\dot{b}}=- | p ]_a \langle p \vert_{\dot{b}}$ and $p^{\dot{a}b} = - \vert p \rangle^{\dot{a}} [ p |^b$. The correspondence between the two-dimensional helicity spinors and four-component Dirac spinors associated with Feynman rules is demonstrated in  Appendix~\ref{sec: DiracHelicity}.

To facilitate and implement such calculations on a GQC, we use \emph{qubits}, the quantum analogue of the \emph{bit} for classical computation. The state of the qubit is defined on a two-dimensional complex vector space with states $\vert 0 \rangle$ and $\vert 1 \rangle$ forming the orthonormal basis for this space. A qubit can thus be formed by a linear superposition of these orthonormal basis states. By considering a general qubit parametrized by two angles 
\begin{align}
\vert  \psi  \rangle=\cos\frac{\theta }{2}\vert  0  \rangle+e^{i\varphi }\sin\frac{\theta }{2}\vert  1  \rangle=\begin{pmatrix}
\cos\frac{\theta }{2}\\ 
\sin\frac{\theta }{2}e^{i\phi }
\end{pmatrix},
\label{eq:qubit}
\end{align}
we can represent the qubit on a three-dimensional unit sphere called the Bloch sphere. Performing unitary operations on qubit states corresponds to rotating states in the Bloch sphere.

Remarkably, comparing Eqs.~(\ref{eq:spinor}) and (\ref{eq:qubit}), helicity spinors can be represented through a qubit, modulo an overall normalisation factor $\sqrt{2E}$, and the calculation of helicity amplitudes follows the identical structure shown in Fig.~\ref{fig:encoding}, i.e. quantum operators act on an initial state to eventually perform the projection onto a final state. 
In contrast to classical computers, where all numerical quantities are converted into a binary system representation, on which an algorithm is applied, and then transformed back into quantities that can be understood in terms of a numerical result, in a quantum computing algorithm, the helicity spinor is a faithful representation of the object the circuit directly operates on. The spinors can be directly represented as vectors on the Bloch sphere, which provides the most efficient encoding of the state on which the algorithm operates.
This indicates that GQC provide an ideal framework for the calculation of helicity amplitudes.

\begin{figure}[ht!]
\centering
\begin{subfigure}{0.24\textwidth}
\centering
\includegraphics[scale = 0.4]{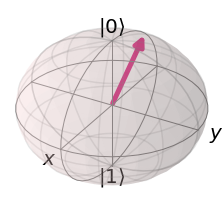}
\subcaption{$\vert p \rangle^{\dot{a}}$}
\end{subfigure}
\begin{subfigure}{0.24\textwidth}
\centering
\includegraphics[scale = 0.4]{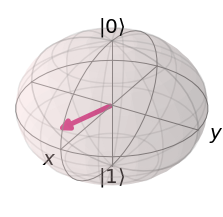}
\subcaption{$\vert p ]_a$}
\end{subfigure}
\begin{subfigure}{0.24\textwidth}
\centering
\includegraphics[scale = 0.4]{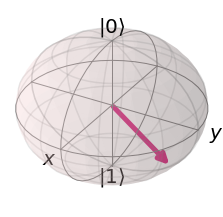}
\subcaption{$(\langle p \vert_{\dot{a}})^\textrm{T}$}
\end{subfigure}
\begin{subfigure}{0.24\textwidth}
\centering
\includegraphics[scale = 0.4]{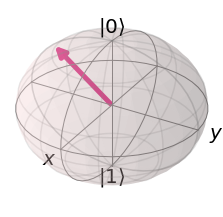}
\subcaption{$([p \vert^{a})^\textrm{T}$}
\end{subfigure}                          
\caption{A visualisation of the helicity spinors $\vert  p  \rangle^{\dot{a}}$,$\langle p \vert_{\dot{a}}$, $(| p ]_a)^\textrm{T}$ and $([ p |^a)^\textrm{T}$ for $\theta=\pi/4$, $\phi=\pi/2$, $E=1/2$ on the Bloch sphere, following the choice of representation of Eq.~(\ref{eq:spinor}).}
\label{fig: bloch}
\end{figure}

Consequently, we will exploit that the spinors used to calculate helicity amplitudes naturally live in the same representation space as qubits. This motivates the manipulation of the direct correspondence of the $\theta$ and $\phi$ variables of the qubit states and helicity spinors to represent the spinors on a quantum circuit. We further encode operators acting on spinors as quantum circuits of unitary operations. These can be applied to qubits (rotating vectors on the Bloch sphere) to calculate helicity amplitudes. The helicity spinors $\vert  p  \rangle^{\dot{a}}$,$(\langle p \vert_{\dot{a}})^\textrm{T}$, $| p ]_a$ and $([ p |^a)^\textrm{T}$ are visualised for $\theta=\pi/4$, $\phi=\pi/2$, $E=1/2$, as vectors on the Bloch sphere in Fig.~\ref{fig: bloch}, in direct analogy to their respective qubit representation. 

This study aims to create the basic building blocks to encode spinor helicity calculations on a quantum circuit. 
 These basic building blocks are then used to construct quantum algorithms for two simple examples of helicity calculations: i) the contraction of an external polarisation vector corresponding to a $g\rightarrow q\bar{q}$ vertex and ii) the construction of s and t-channel amplitudes for a $q\bar{q}\rightarrow q\bar{q}$ process with identical initial and final quark flavours. `Helicity registers' are crucially introduced into these circuits to control the helicity of each particle involved. In addition, we introduce a superposition state between the helicity qubits of $\vert + \rangle =\vert 1\rangle$ and $\vert - \rangle =\vert 0 \rangle$ by applying Hadamard gates to the helicity registers. In doing so, we can calculate both helicities of each particle involved simultaneously, thus fully utilising the quantum nature of the computation. This advantage is further exploited by the simultaneous computation of s and t-channel amplitudes for the $q\bar{q}\rightarrow q\bar{q}$ process. 
 
This section is organised as follows: a description of the quantum circuit for the 1$\rightarrow$2 process of $g\rightarrow q\bar{q}$ is given in Sec.~\ref{sec:vertex}, together with a comparison of the results of the algorithm as run on a real machine and a simulator, the quantum circuit and the results for the 2$\rightarrow$2 process of $q\bar{q}\rightarrow q\bar{q}$ are given in Sec.~\ref{sec:qqqq}, and a brief discussion of the generalisation of the algorithm to 2$\rightarrow n$ processes follows in Sec.~\ref{sec: helicityFuture}.

\subsection{Constructing helicity spinors and scalar products on the Bloch sphere}

The helicity spinors have been implemented on the quantum circuit by constructing Bloch sphere representations, like the ones shown in Fig.~\ref{fig: bloch}. The helicity spinor decompositions are outlined in detail in Appendix~\ref{sec: helicityAmpGates}. They utilise the Qiskit $U_3 (\theta, \phi, \lambda)$ gate, which applies a rotation to a single qubit. The rotation is defined by,

\begin{equation}
U_3 (\theta, \phi, \lambda) =  \begin{pmatrix} \cos \big( \frac{\theta}{2} \big) & - e^{i\lambda} \sin \big( \frac{\theta}{2} \big) \\ e^{i\phi} \sin \big( \frac{\theta}{2} \big) & e^{i(\phi + \lambda)} \cos \big( \frac{\theta}{2} \big) \end{pmatrix}.
\end{equation}
A simple $U_3$ gate acting on a $\vert 0 \rangle$ state has been used to create the $\vert q \rangle^{\dot{a}}$ spinor, where $\theta $ and $\phi $ variables of the $U_3$ gate corresponded to the $\theta $ and $\phi $ variables of the helicity spinor. The $\vert q ]_a$ spinor has been created by sequentially applying a $U_3^\dagger$ rotation and a $NOT$ gate, where here the $\theta $ and $ \lambda  $ variables of the $U_3$ gate corresponded to the $\theta $ and $\phi $ variables  of the $\vert q ]_a$ spinor. 

To construct the scalar products $\langle pq \rangle$ or $[pq]$ on a quantum computer, $2\times2$ unitary gates $U_{\langle p}$ and $U_{[p}$ were created such that, when they act on the $\vert q \rangle^{\dot{a}}$ and $\vert q]_a$ spinors respectively, the scalar product values correspond to the first component of the final qubit state, i.e. the complex coefficient associated with the $\vert 0 \rangle$ state. It should be noted that the factors of $\sqrt{2E}$ in the definition of the helicity spinors have not been accounted for such that the spinor-qubit states are normalized to one on the quantum register. As a consequence, these factors must be added after the results have been obtained from the quantum computer.

\subsection{1$\rightarrow$2 amplitude calculation}
\label{sec:vertex}
\begin{figure}[ht!]
\centering
\includegraphics[scale = 0.5]{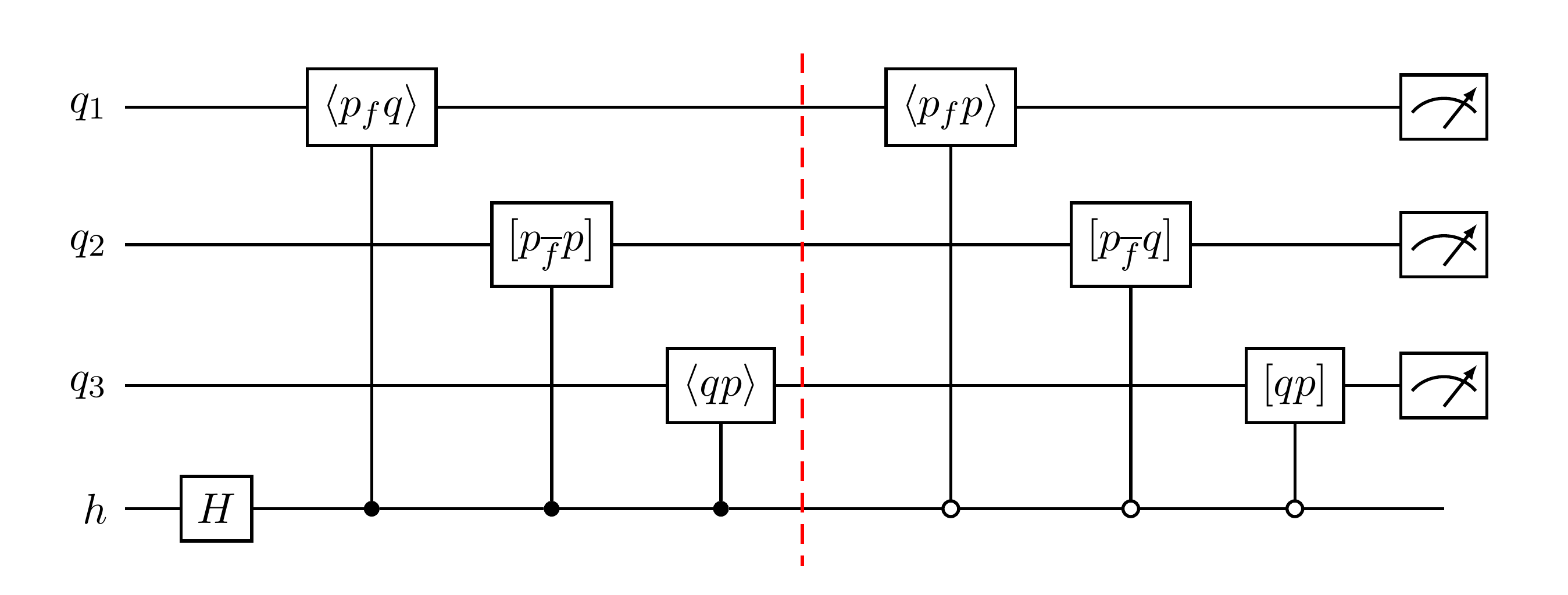}
\caption{$gq\overline{q}$ vertex circuit. The amplitude for the process is calculated on the $q_i$ qubits, which are controlled from the helicity register. The $q_i$ qubits are then measured by the quantum computer.}
\label{fig: gqqbarCircuit} 
\end{figure}

A simple application of the helicity amplitude approach is the calculation of a 1$\rightarrow$2 process. Here we will consider the process of $q \rightarrow g\overline{q}$ by calculating the $gq\overline{q}$ vertex,

\begin{align}
\mathcal{M}_{gq\overline{q}} &=  \langle p_f \vert \bar{\sigma }_{\mu } \vert p_{\overline{f}} ] \epsilon^\mu_\pm,
\end{align}
where $p_f$ and $p_{\overline{f}}$ are the momenta associated with the fermon and anti-fermion respectively. The gluon polarisation vectors are defined as~\cite{elvang_huang_2015},

\begin{align}
\epsilon^\mu_+ &= - \frac{\langle q \vert \bar{\sigma }^{\mu } \vert p  ]}{\sqrt{2} \langle q p \rangle}, &\epsilon^\mu_- = - \frac{\langle p \vert \bar{\sigma }^{\mu } \vert q ]}{\sqrt{2} [q p]}.
\label{eq:polarisation}
\end{align}
From this, it is possible to create a circuit where each four-vector present in the amplitude, i.e. the fermion anti-fermion vertex and polarisation vector, is calculated individually on a series of 4 qubits. This is done by using the corresponding Pauli gates for each four-vector component on each qubit. However, this will lead to a large circuit depth due to the number of gates required to do such a calculation. Therefore it is useful to simplify the expression for the amplitude using the Fierz identity,

\begin{equation}\label{eqn: Fierz}
\langle p \vert \bar{\sigma }^{\mu } \vert q ] \langle k \vert \bar{\sigma }_{\mu }\vert l] = 2 \langle p k \rangle [q  l ].
\end{equation}
With this, the amplitude for the $gq\overline{q}$ vertex becomes

\begin{align}\label{eqn: gqqbarAmp}
\mathcal{M}_+ &= -\sqrt{2} \frac{\langle p_{f} q \rangle [ p_{\overline{f}} p ]}{ \langle q p \rangle}, &\mathcal{M}_- = - \sqrt{2} \frac{\langle p_f p \rangle [p_{\overline{f}} q]}{ [q p]}.
\end{align}
As a consequence of this simplification, the number of qubits needed to calculate the amplitude on the quantum computer can be reduced from 10 to 4. The circuit for calculating this amplitude is shown in Fig.~\ref{fig: gqqbarCircuit}. The three $q_i$ qubits calculate the three scalar products from Eq.~(\ref{eqn: gqqbarAmp}) using the gate decompositions outlined in Appendix~\ref{sec: helicityAmpGates}. These rotation gates are controlled from the helicity register, $h$. If $h$ is in the $\vert 1 \rangle$ state, then the helicity is positive and the $\mathcal{M}_+$ amplitude is calculated; if $h$ is in the $\vert 0 \rangle$ state, then the helicity is negative and the $\mathcal{M}_-$ amplitude is calculated. The three calculation qubits, $q_i$, are then measured by the quantum machine.

\begin{figure}[ht!]
\centering
\begin{subfigure}{\textwidth}
\centering
\includegraphics[scale = 0.4]{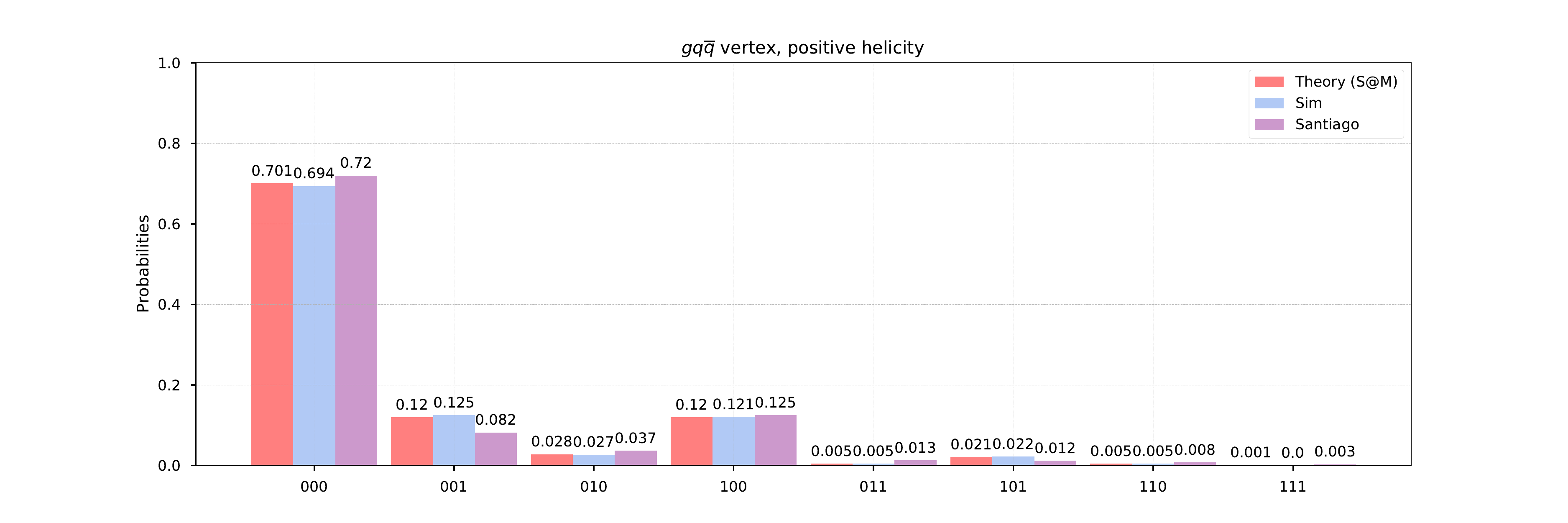}
\end{subfigure}
\begin{subfigure}{\textwidth}
\centering
\includegraphics[scale=0.4]{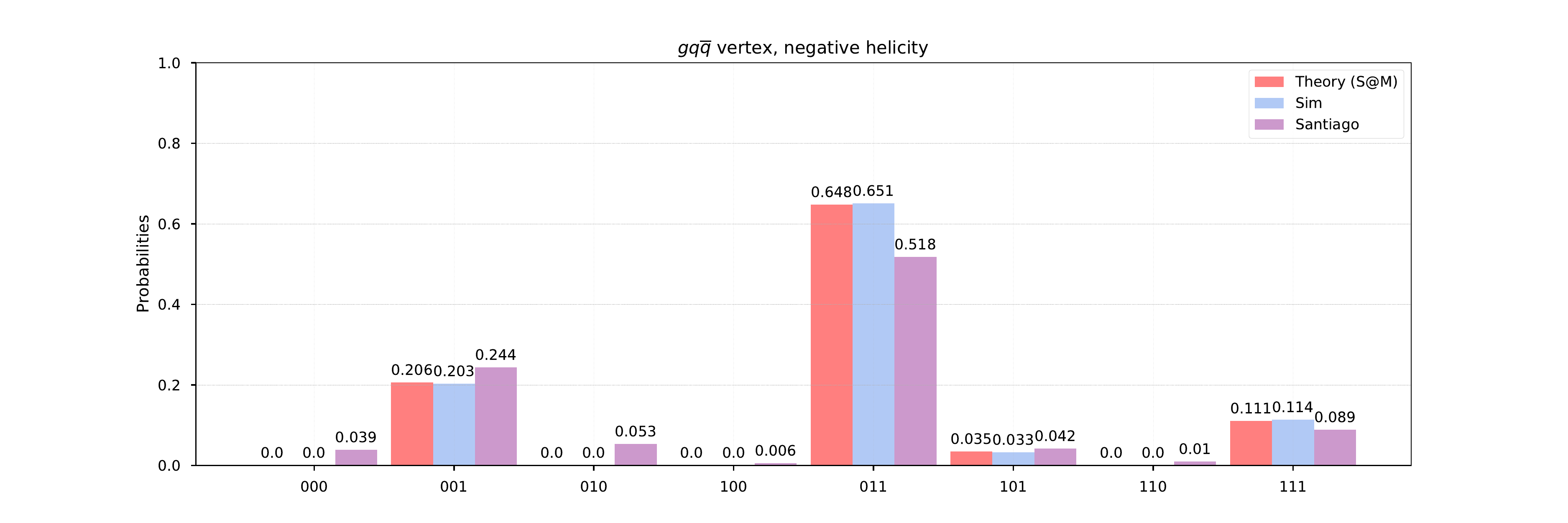}
\end{subfigure}                      
\caption{Results for the $q\rightarrow g\overline{q}$ helicity amplitude calculation. Comparison between theoretically calculated probability distribution, quantum simulator and real quantum computer.}
\label{fig:gqqbarResults}
\end{figure}

Figure~\ref{fig:gqqbarResults} shows the results of the algorithm for a random selection of small scattering angles, with runs on the IBM Q 32-qubit Quantum Simulator~\cite{32_sim} and the IBM Q 5-qubit Santiago Quantum Computer~\cite{ibmq_santiago}; both of which have been compared to theoretical predictions of the probability distributions extrapolated directly from analytic calculations of the helicity amplitude, calculated using the S@M software~\cite{zbMATH05804587}. The simulator has been run without a noise profile for 10,000 shots. The results agree well with theoretically predicted values, to within 1$\sigma$. From these distributions, one can determine the helicity setup of the process and consequently reconstruct the helicity amplitudes.

The Santiago machine has been run on the maximum shot setting of 8192 for 100 runs, leading to a total of 819,200 shots of the algorithm. Figure~\ref{fig:gqqbarResults} shows that the quantum computer's performance does not match that of a perfect machine, as expected. Therefore, the simulator is rerun with the noise profile of the Santiago device and a comparison between this and the quantum computer is shown and discussed in Appendix~\ref{app:helicityAmpCalc}. 

The results from the quantum computer, shown in Fig.~\ref{fig:gqqbarResults}, have been achieved by isolating the individual helicity processes on the quantum circuit, and removing the superposition between the positive and negative processes. The full amplitude is achieved through the implementation of a Hadamard gate on the helicity qubit, which puts the system into a superposition state of the positive and negative processes. The helicity of the process is then determined by measuring the helicity register. The qubit setup chosen here has been used in order to best reduce the \textit{CNOT} qubit errors and limits the number of \textit{SWAP} operations needed in the algorithm. The Santiago machine is a 5-qubit quantum computer, with all qubits connected inline to their adjacent qubit. The helicity qubit, $h$, from Fig.~\ref{fig: gqqbarCircuit} has been assigned to qubit 4 on the Santiago machine, with the $q_i$ qubits on the 2nd, 3rd and 5th qubits of the Santiago machine. The optimum qubit setup would have the $h$ qubit fully connected to the $q_i$ qubits, thus fully minimising the \textit{SWAP} operation errors. However, the available machines with such a qubit mapping on the public IBM Q experience have a lower quantum volume than the Santiago machine, which reports a quantum volume of 32. Consequently, the trade of ideal qubit mapping for a better quantum volume has been made. 

\begin{figure}[ht!]
\centering
\includegraphics[width = 0.5\textwidth]{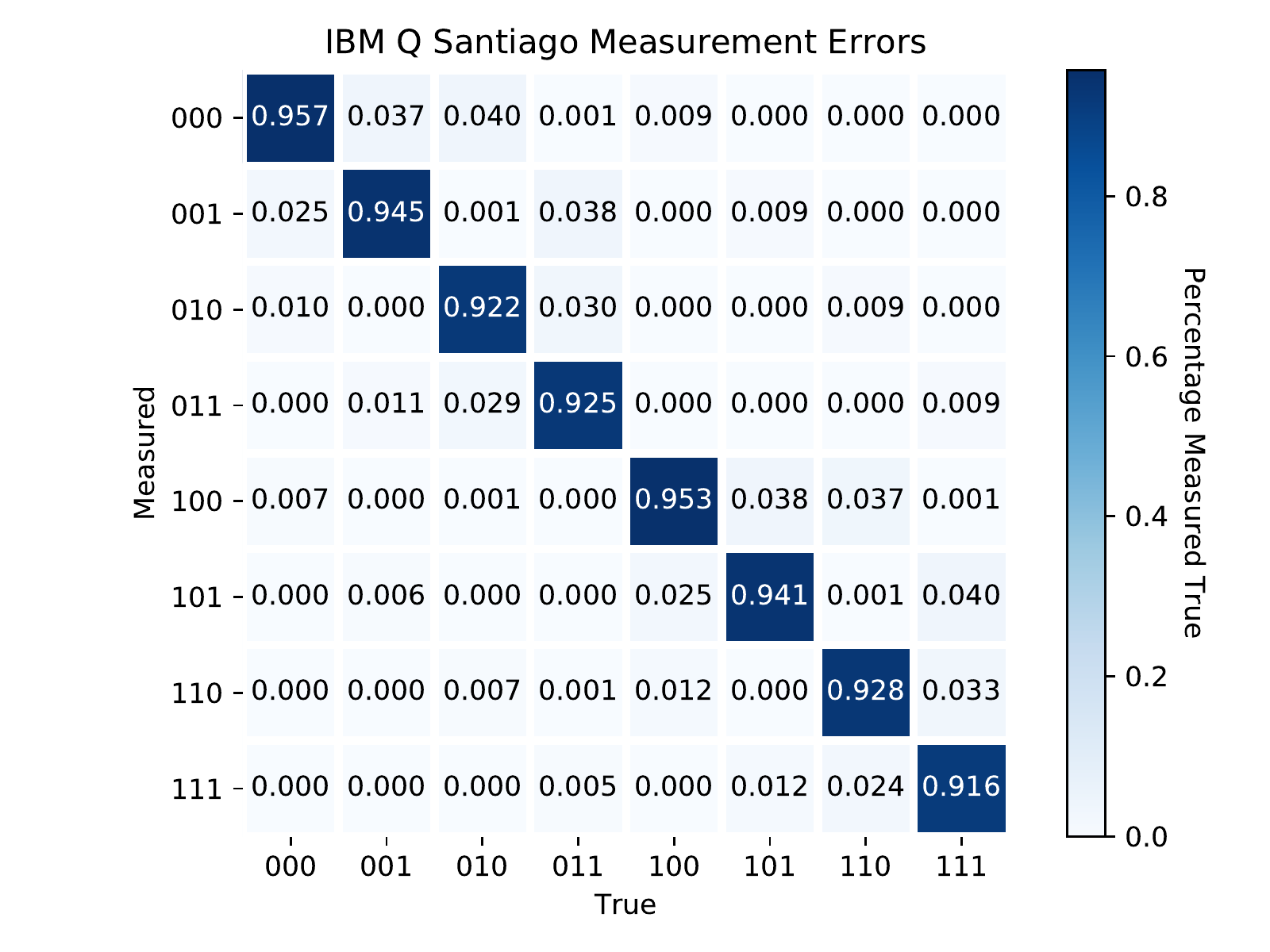}
\caption{IBM Q Santiago 5-qubit Quantum Computer Response Matrix for measurement error correction on the 4 qubit helicity amplitude calculation algorithm.}
\label{fig:responseMatrix}
\end{figure}

One of the key sources of error in the quantum computer is readout noise. Error mitigation methods have been used to optimise the output from the quantum computer and reduce readout noise effects. This has been done using the Qiskit Ignis software~\cite{IBMQ}, which provides tools for noise characterisation and error correction based on noise models of the quantum machines. The method involves testing simple qubit states on a series of calibration circuits, which are run using the quantum simulator with the noise profile of the Santiago machine. The response matrix created from this is shown in Fig.~\ref{fig:responseMatrix}. This response matrix is calculated immediately before running the algorithm and then applied to the machine results to obtain the error corrected results, as shown in Fig.~\ref{fig:gqqbarResults}. 

\subsection{2$\rightarrow$2 amplitude calculation}
\label{sec:qqqq}

Extending from the $1\rightarrow 2$ case in Sec.~\ref{sec:vertex}, the implementation of a full helicity amplitude calculation for the s and t-channels of a $2 \rightarrow 2$ scattering process is presented here\footnote{Note, for the calculation of the $1 \to 3$ case only minor modifications are needed.}. As an example, we consider a $q\overline{q} \rightarrow q\overline{q}$ process. The initial state quark and antiquark are labelled as particles 1 and 2 respectively and the final state quark and antiquark as 3 and 4. In total, there are only 4 non-zero helicity configurations possible for each s and t-channel process. The relevant amplitudes are,

\begin{align}\label{eqn: 2to2ampsfull}
\mathcal{M}_{s{(+-+-)}} &= -\langle 2 \vert \bar{\sigma }^{\mu } \vert 1] \frac{1}{s_{12}}[ 3 \vert \sigma _{\mu } \vert 4 \rangle, &\mathcal{M}_{s{(+--+)}} =  -\langle 2 \vert \bar{\sigma }^{\mu } \vert 1] \frac{1}{s_{12}}\langle 3 \vert \bar{\sigma }_{\mu } \vert 4 ]
\end{align}
and 
\begin{align}\label{eqn: 2to2amptfull}
\mathcal{M}_{t{(++--)}} &= -\langle 3 \vert \bar{\sigma }^{\mu } \vert 1] \frac{1}{s_{13}}[ 2 \vert \sigma _{\mu } \vert 4 \rangle, &\mathcal{M}_{t{(+--+)}} =  -\langle 3 \vert \bar{\sigma }^{\mu } \vert 1] \frac{1}{s_{13}}\langle 2 \vert \bar{\sigma }_{\mu }\vert 4 ]
\end{align}
where the +/- signs denote the helicity of the outgoing-particles 1, 2, 3 and 4 and

\begin{equation}
s_{ij} = -(p_{i}+p_{j})^{2}=\langle i j \rangle [j i ].
\end{equation}
The other non-zero amplitudes are obtained by complex conjugation.

The calculation is performed in the Centre-of-Mass (CM) frame and the momenta of individual particles is defined such that the only dependent input variable is the angle, $\theta$, through which the quark (and antiquark) is scattered. In the CM frame, the overall magnitude of energy, $E$, associated with the momenta of each particle also drops out of the final helicity amplitude and is therefore not considered in this example.

\begin{figure}[ht!]
\centering
\includegraphics[scale = 0.5]{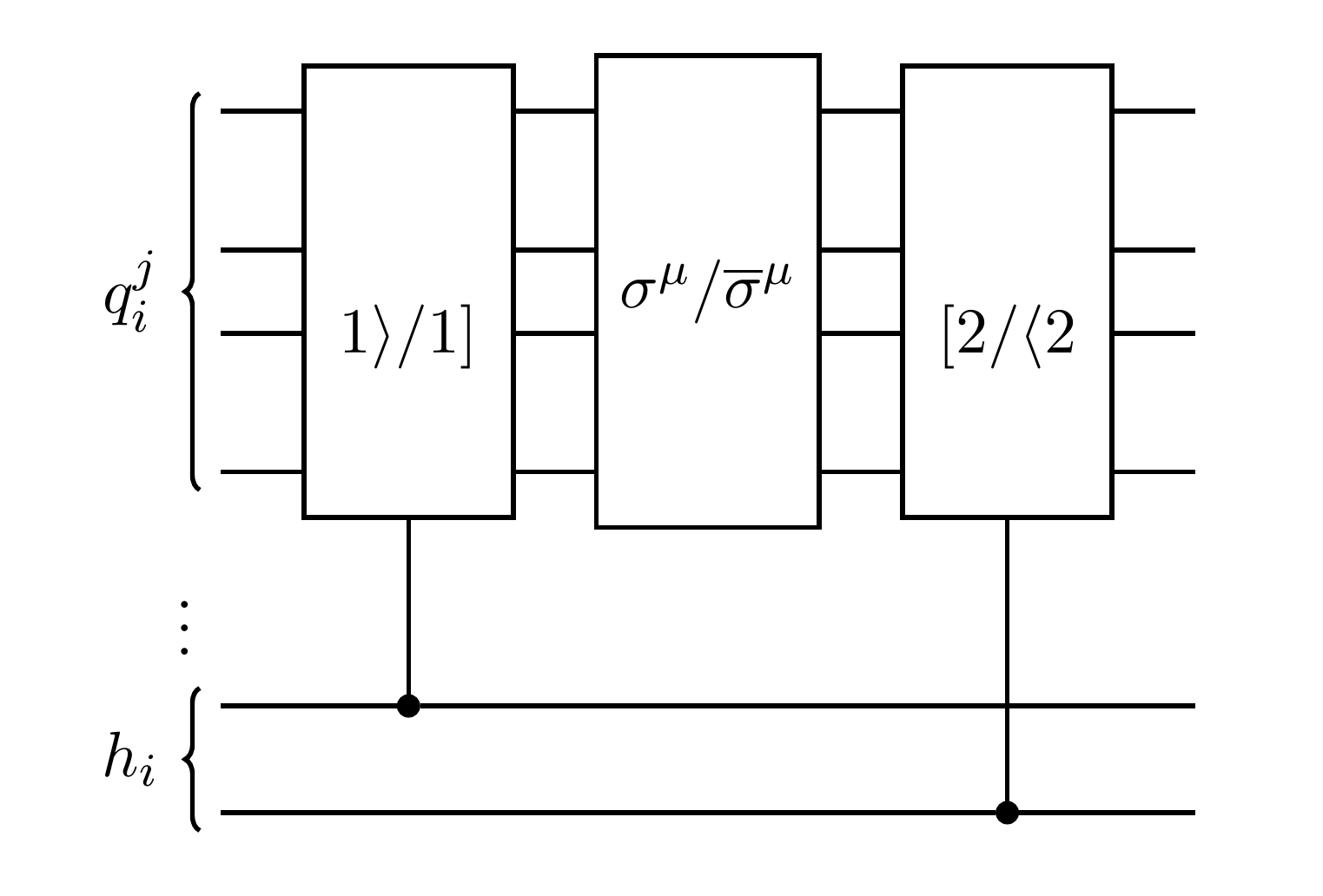}
\caption{Circuit for the $q\overline{q}\rightarrow q\overline{q}$ process helicity amplitude calculation. The $q^j_i$ registers are used to calculate the $q\overline{q}$ vertices, and these are controlled from the helicity registers, $h_i$, which dictate the helicity configuration of the process.}
\label{fig: 2to2}
\end{figure}

In the `all-outgoing' convention of spinor-helicity formalism~\cite{elvang_huang_2015}, the momenta of incoming particles are flipped so that the incoming quark (1) (antiquark (2)) is mapped to an outgoing antiquark (quark) with opposite helicity. In the quantum algorithm, each quark-antiquark vertex is calculated on a 4-qubit quantum register, $q_i$. The outgoing antifermion spinor, $ q \rangle/q]$, is implemented on the vertex quantum register, $q_i^j$, followed by the two dimensional representation of the gamma matrices, $\sigma ^{\mu }/\bar{\sigma }^{\mu }$, and then finally the vertex is closed with the opposite helicity outgoing fermion spinor, $[q/\langle q$. A single qubit, $s$, is used to calculate the denominator of the gluon propagator. The calculation is controlled both from the helicity registers, $h_i$, which determine what helicity configuration the particles are in, and the amplitude qubit, $p$, which controls whether the s or t-channel process is calculated. A schematic of the quantum circuit is shown in Fig.~\ref{fig: 2to2}. Through this implementation, each component of the helicity amplitude can be calculated and extracted from the machine.

This method is powerful as it allows for each component of the calculation to be extracted, however it leads to a complicated circuit, especially if one implements a method of dealing with incorrect helicity setups.  As in Sec.~\ref{sec:vertex}, the circuit can be simplified by directly calculating the scalar products required for the final amplitudes. The amplitudes given in Eqs.~(\ref{eqn: 2to2ampsfull}) and~(\ref{eqn: 2to2amptfull}) can be simplified using Eq.~(\ref{eqn: Fierz}) (and that $ [p \vert \sigma ^{\mu } \vert q \rangle = \langle q \vert \bar{\sigma }^{\mu } \vert p])$ to give the final forms,

\begin{align}\label{eqn: 2to2amps}
\mathcal{M}_{s_{(+-+-)}} &= 2\frac{\langle 2 4 \rangle [3 1]}{\langle 1 2 \rangle [2 1]}, &\mathcal{M}_{s_{(+--+)}} =  2\frac{\langle 2 3 \rangle [4 1]}{\langle 1 2 \rangle [2 1]}
\end{align}
and 
\begin{align}\label{eqn: 2to2ampt}
\mathcal{M}_{t_{(++--)}} &= 2\frac{\langle 3 4 \rangle [2 1]}{\langle 1 3 \rangle [3 1]}, &\mathcal{M}_{t_{(+--+)}} =  2\frac{\langle 3 2 \rangle [4 1]}{\langle 1 3 \rangle [3 1]}.
\end{align}
Using these expressions, the number of qubits needed for the circuit is reduced from 17 to 12 qubits. Another advantage is that the machine now only has to read out 3 qubits, where previously 8 qubits were read out per run. On these three qubits, each of the scalar products is calculated. The quark-antiquark vertex scalar products from the numerator are calculated on the first two qubits, and the denominator of the gluon propagator is calculated on the third qubit. Only one scalar product needs to be calculated for the denominator since~\cite{elvang_huang_2015},

\begin{equation}
\langle i j \rangle = [ j i ] ^*,
\end{equation}
therefore the second scalar product can be determined from the same qubit. 

\begin{figure}[!h]
\centering
\includegraphics[width=\textwidth]{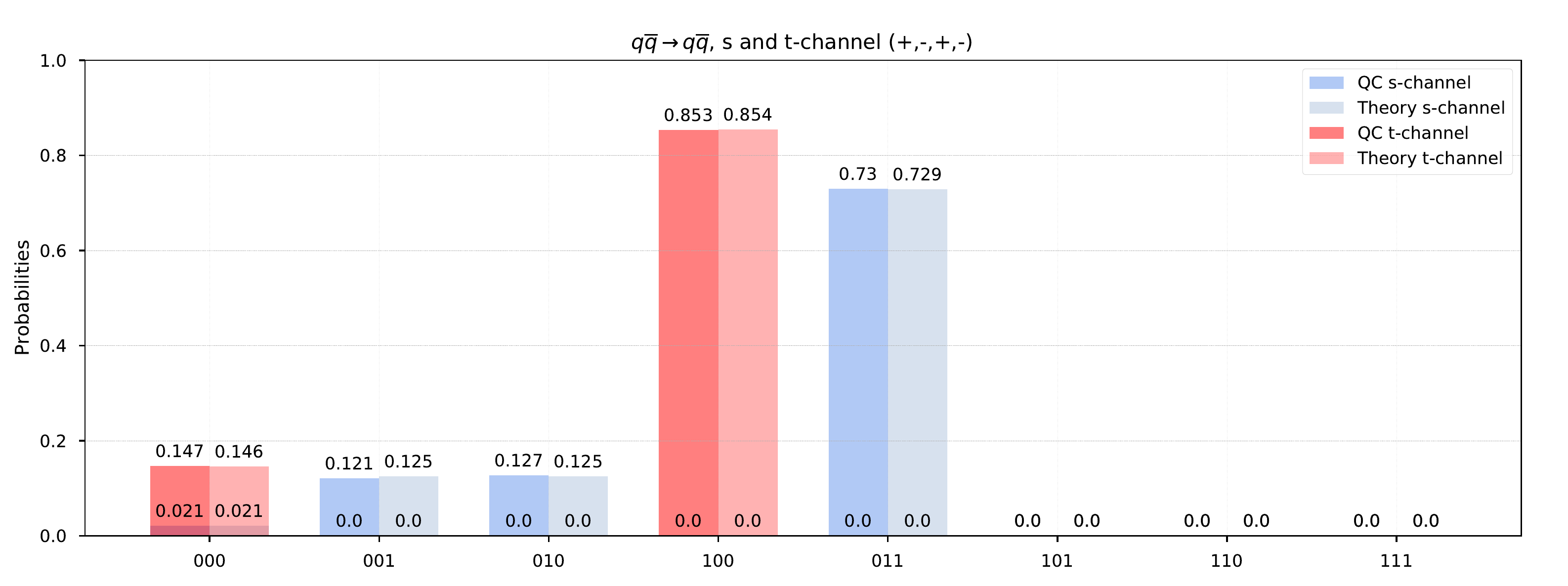}
\caption{Comparison between theoretically predicted qubit final state probabilities and 32-qubit quantum simulator output for the s and t-channel $q\overline{q} \rightarrow q\overline{q}$ process in the (+,-,+,-) helicity configuration. The quark (antiquark) scattering angle has been chosen as $\theta_3$ = $\frac{\pi}{4}$.}
\label{fig: 2to2Comparison}
\end{figure}

This simplified circuit is run on the IBM Q 32-qubit Quantum Simulator~\cite{32_sim} for 10,000 runs and compared to theoretically calculated probability distributions, extrapolated directly from analytic calculations of the helicity amplitude, calculated using the S@M software~\cite{zbMATH05804587}. Using the equivalence between helicity spinors and orthogonal pure state qubits, these theoretical predictions have been obtained from the probabilities of each of the qubits to be in the $\vert 0 \rangle$ or $\vert 1 \rangle$ state, which correspond to the magnitude squared of the upper and lower components of the helicity spinor respectively. The results from the quantum simulator show that the output of the quantum circuit lies within 1$\sigma$ of the theoretically predicted probability distribution and are shown in Fig.~\ref{fig: 2to2Comparison} for both the s and t-channel in a specific helicity configuration. 

\subsection{Generalisation to $2 \to n$ amplitude calculations}\label{sec: helicityFuture}

It can be shown, using the BCFW recursion formula \cite{Cachazo:2004kj, Britto:2005fq} and the relations in Eq.~(\ref{eq:polarisation}), that scattering amplitudes for massless partons can be reduced to a combination of scalar products between helicity spinors\footnote{A well-known example is the Parke-Taylor formula for a $2\to n$ gluon scattering process, where the gluons i and j have helicity (-) and all other gluons have helicity (+). Then the formula provides the following expression for the amplitude $\mathcal{A}_n$,
\begin{equation}
\mathcal{A}_n[1^+ \cdots i^- \cdots j^- \cdots n^+]= (-g_s)^{n-2} \frac{\left < ij \right >^4 }{\left < 12 \right > \left < 23\right > \cdots \left < n1 \right >}.
\end{equation}
}. Consequently, the algorithm presented in Secs.~\ref{sec:vertex} and \ref{sec:qqqq} can be generalised to multi-particle amplitudes straightforwardly, as the tools are already created, namely the circuit decompositions of the helicity spinors from Appendix~\ref{sec: helicityAmpGates}. The number of calculation qubits, $q_i$, and the number of helicity qubits, $h_i$, needed in the algorithm both scale linearly with the number of final state particles, $n$. As the number of helicity qubits, $h_i$, scales linearly, then so does the number of work qubits needed in the algorithm. Each scalar product calculation requires two spinor operations, and so the algorithm can be easily extended without adding disproportionate complexity. The circuit depth scales linearly with an increase in the number of scalar products, calculated on the $q_i$ qubits, and the number of helicity qubits, $h_i$, added to the circuit. 

It is interesting and practical to consider the extension of the simple helicity amplitude algorithms presented here to more complicated processes that are likely to be present in high energy collisions, such as those studied at the LHC. As we have seen in Sec.~\ref{sec:vertex}, modern public access quantum computers do not perform to a standard where one could extrapolate accurate calculations of helicity amplitudes, even for a single vertex. However, the performance of public access computers is well below that of state of the art machines, such as the IBM 53-qubit machine and the Honeywell machine. The latter, in unpublished work, claims to have the world's best Quantum Volume of 64~\cite{yirka_2020}. Such computers do not have the same restrictions as the smaller, less capable public access machines. The more powerful machines offer more choice for qubit setup and mapping, and the ability to perform more operations before decoherence in the machine starts to affect the circuit output. We can speculate that the algorithms presented here would be very accurate on these machines, especially the vertex calculation, which comprises a maximum of only 33 operations across 4 qubits.

The main difficulty of extending such algorithms for helicity amplitude calculations on quantum computers comes not only from limitations due to the number of qubits, but also the machine's fault tolerance. The more complicated the helicity amplitude calculation, the more operations are needed to calculate it. Therefore, a machine needs not only sufficient qubits but also the ability to implement many operations without excess noise. For the algorithm proposed, the immediate challenge is not the number of qubits available, but the number of operations that can be reliably implemented on the circuit. With advancements in the Quantum Volume of quantum computers~\cite{Jurcevic2020DemonstrationOQ}, this limitation will likely be overcome on current hardware. It is possible that near-future computers will have the ability to perform accurate and precise calculations and also have a large number of qubits. IBM recently announced their roadmap for the future and the goal of having machines with the number of qubits exceeding 1,000 by 2023~\cite{ibm_future}. Therefore, it is highly likely that these near-future devices will be able to perform precise helicity amplitude calculations for processes with a large number of particles.

\section{Parton shower algorithm}
\label{sec:parton}

After the hard process is calculated, the next step in simulating a scattering event at a high-energy collider experiment is the parton shower stage. The parton shower evolves the scattering process from the hard interaction scale down to the hadronisation scale. We propose an algorithm for simulating a QCD parton shower using IBM Quantum Experience \cite{IBMQ} software and hardware. The quantum circuit has been implemented to simulate a 2-step QCD parton shower with collinear splittings only. Section~\ref{sec: showertheory} provides the theoretical outline for the shower algorithm and discusses the splitting functions and probability calculations implemented in the quantum circuit. A brief overview of the quantum circuit is given in Sec.~\ref{sec: showerimplementation}, and a comparison between the results of the algorithm and theoretically calculated probability distributions are discussed in Sec.~\ref{sec: showerresults}. A glossary of quantum logic gates is given in Appendix~\ref{app: definitions} and a detailed overview of the quantum circuit for the algorithm in Appendix~\ref{app: showercircuit}.

\subsection{Theoretical outline of shower algorithm}\label{sec: showertheory}

We present a parton shower algorithm with the ability to simulate a general, discrete QCD parton shower, harnessing the quantum computer’s ability to remain in a quantum state throughout the algorithm. In contrast to classical methods, the algorithm does not need to explicitly keep track of individual shower histories. Instead, our algorithm constructs and maintains a wavefunction that consists of a superposition of all possible shower histories, with the final measurement projecting out a specific quantity of the final state. Consequently, the algorithm presented inherently simulates the quantum interference between all possible final states, without the need for extensive computational logic present in current classical algorithms. In a classical algorithm, a physically meaningful quantity can only be extracted from a parton shower calculation after summing over all possible shower histories, requiring them to be stored on a physical memory device. Our quantum algorithm avoids the need for such an intermediate step, as the measurement is performed on the superposition of all shower histories directly.

The goal is to create the foundation for constructing a general quantum algorithm that can simulate a full QCD parton shower. To comply with the current capabilities of public access quantum computers and simulators provided by IBM Quantum Experience \cite{IBMQ}, the algorithm presented here uses a simplified model consisting of one flavour of quark and a gluon. This reduces the number of qubits needed, and the algorithm can be run on the IBM Q 32-qubit Quantum Simulator \cite{32_sim}. To further reduce the number of required qubits, only collinear splittings are considered within the model. By neglecting the soft-limit, there is no need to keep track of the detailed kinematics of the particles in the shower history.

Collinear emission occurs when a parton splits into two massless particles which have parallel 4-momenta, such that the total momentum, $P$, is distributed between the particles as
\begin{align}\label{Eqn. CollinearEmission}
p_i &= x P, &p_j = (1- x) P,
\end{align}
thus, $(p_i + p_j)^2 = P^2 = 0$ \cite{Taylor_2017}.

The emission probabilities in the algorithm are calculated using the collinear splitting functions outlined in~\cite{Dokshitzer:1977sg,Gribov:1972ri,ALTARELLI1977298,Marzani_2019}. A consequence of the collinear limit being a semi-classical interpretation with 1-to-2 splittings leads to the presence of a diagonal colour charge in the splitting functions, $C_{ii}$. The splitting for a quark to a gluon and a quark, with momentum fractions $z$ and $1-z$ respectively, is described at Leading Order (LO) by

\begin{equation}\label{eqn: quarkSplittings}
P_{q\rightarrow qg}(z) = C_F \frac{1 + (1 - z)^2}{z},
\end{equation}
with $C_F = 4/3$. The gluon splitting can be divided into two parts, with the first describing the splitting of a gluon to a quark-antiquark pair and the second describing the splitting of a gluon to two gluons,

\begin{align}\label{eqn: gluonSplittings}
P_{g \rightarrow q\overline{q}}(z) = n_f T_R (z^2+ (1-z)^2), & &P_{g\rightarrow gg}(z) = C_A \Big[  2 \frac{1 - z}{z} + z(1-z) \Big],
\end{align}
where $C_A=3$ and $T_R=1/2$.
Here, $n_f$ is the number of massless quark flavours, and $T_R$ is the colour factor. It should be noted that both splitting functions have a soft singularity at $z$ = 0; the hard-collinear limit only takes into account finite $z$.

Further to calculating the splitting functions, the Sudakov factors have been used to determine whether an emission occurred in the step. The Sudakov factors for a QCD process are given by \cite{Sudakov:1954sw}

\begin{equation}
\Delta_{i, k} (z_1 , z_2)= \exp \Big[ - \alpha_s^2 \int^{z_2}_{z_1} P_k(z^\prime) dz^\prime \Big],
\end{equation}
and are used to calculate the non-emission probability. The running of the strong coupling, $\alpha_s$, is not simulated in this algorithm and for ease has been set to 1. For any given step $N$, there are $N$ possible particles present, and so the probability that none of the particles split is given by

\begin{equation}\label{eqn: sudakovs}
\Delta_{\textrm{tot}} (z_1, z_2) = \Delta_g^{n_g} (z_1, z_2) \Delta_q^{n_q}(z_1, z_2) \Delta_{\overline{q}}^{n_{\overline{q}}}(z_1, z_2).
\end{equation}
Finally, the probability of a certain splitting is therefore obtained from 

\begin{equation}\label{Eqn. probCalc}
\textrm{Prob}_{k \rightarrow ij} = \big( 1 - \Delta_k \big) \times P_{k \rightarrow ij} (z).
\end{equation}
To implement the algorithm efficiently, preference has been given to gluons splitting to a quark-antiquark pair. This splitting preference implementation is explained in depth in Appendix~\ref{app: showercircuit}, but, for definiteness, the probability of a gluon splitting to two gluons is calculated as

\begin{equation}\label{eqn: probCalc}
\textrm{Prob}_{g \rightarrow gg} = \big( 1 - \Delta_g \big) \times \big( 1 - P_{g \rightarrow q\overline{q}} (z) \big) \times P_{g \rightarrow gg} (z).
\end{equation}
For the energy scale considered here, this should have a small affect on the results as $P_{g \rightarrow q\overline{q}}(z)~\ll ~P_{g \rightarrow gg}(z)$.

\subsection{Implementation on quantum circuit}\label{sec: showerimplementation}

\begin{figure}[!h]
\centering
\includegraphics[scale = 0.5]{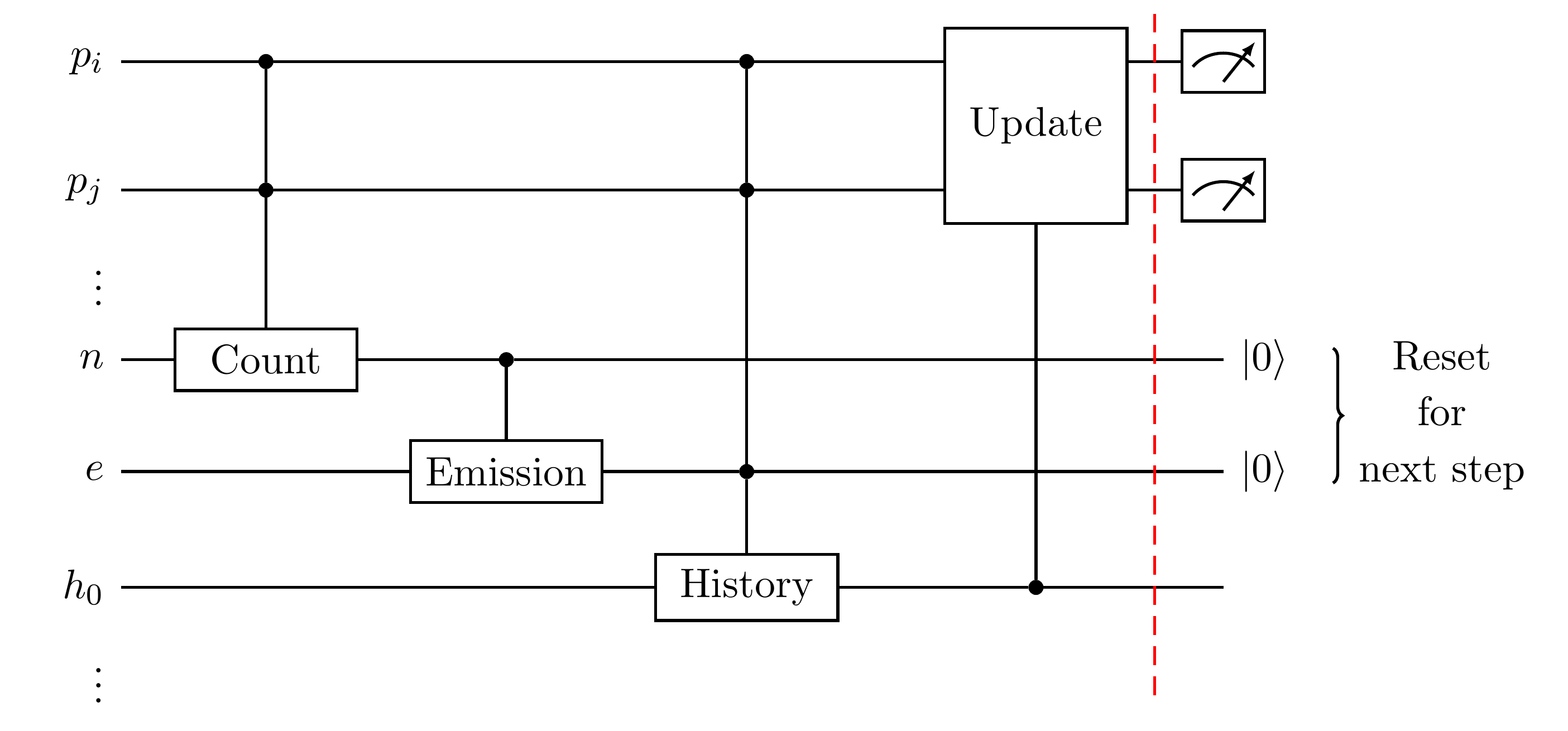}
\caption{Circuit diagram for one step of the algorithm. The circuit comprises particle registers, emission registers and history registers.}
\label{fig: 1stepcircuit}
\end{figure}

A quantum circuit has been constructed to simulate a parton shower with collinear splittings. The circuit comprises of particle registers, emission registers and history registers and uses a total of  31 qubits. The algorithm is discretised into individual steps. An emission can occur in each step, and the probabilities are calculated from the splitting functions and Sudakov factors. To meet the 32 qubit limit of the IBM Q Quantum Simulator \cite{32_sim}, the algorithm has been limited to two steps, but it is generally extendable. Figure~\ref{fig: 1stepcircuit} shows the circuit diagram for a single step.

The algorithm follows a similar method to that described in~\cite{bauer2019quantum}, first counting the particles present in the simulation, determining whether an emission has occurred and if so, assessing which splitting did occur, then finally updating the particle content of the simulation. In contrast to the method shown by \cite{bauer2019quantum}, the algorithm presented here has the ability to simulate a QCD process with splittings for both gluons and quarks implemented using the splitting functions outlined in Eqs.~(\ref{eqn: quarkSplittings}) and (\ref{eqn: gluonSplittings}). The addition of such splitting functions leads to significant changes to the algorithm compared to that presented in~\cite{bauer2019quantum}, specifically in the History and Update gates of the algorithm, shown in Fig.~\ref{fig: 1stepcircuit}. The implementation of these gates is outlined in detail in Appendix~\ref{app: showercircuit}. Unlike the algorithm presented in~\cite{bauer2019quantum}, we have chosen not to introduce flavour mixing at the start of the algorithm. Instead, the superposition and interference between the possible output states are introduced in the tailored History and Update gates. With the ability to simulate gluon and quark splittings, the algorithm is thus well suited to hadronic parton shower simulation and provides the foundations for a general parton shower algorithm for use on a GQC. 

The parton shower algorithm is designed to operate on the public access IBM~Q 32-qubit Quantum Simulator~\cite{32_sim}, which allows for a total of two steps to be simulated on the machine. As the machine is a simulator, it does not suffer from noise or a limit on the number of operations due to qubit decoherence effects, therefore giving a simulation of a perfect machine. As a consequence, error checking is easily done with direct comparison to theoretically predicted probability distributions, and this is discussed in Sec.~\ref{sec: showerresults}. 

One of the main benefits of using a quantum computing (QC) algorithm for the simulation of QCD parton showers over classical methods is the computational simplicity of the algorithm. When dealing with interference of different splittings in the shower process, the algorithm presented here offers a much less computationally complex approach than that provided by modern Monte Carlo event generators. This is achieved by utilising the unique ability to maintain the quantum computer in a fully quantum state throughout the algorithm, and only collapse to a classical circuit by measurement at the end of the process. This allows for the system to account for all possible parton shower histories simultaneously. In contrast, modern Monte Carlo methods must manually keep track of the particle splitting histories to consider all possible contributions to a specific final state. For a two-step, discrete parton shower, this is a relatively easy task for a modern Monte Carlo generator. However, the quantum computing field is still in its infancy; the true potential of quantum computing for simulating QCD parton showers will become apparent with the advancement of quantum technologies. With more available qubits and machines with improved hardware, the algorithm presented here will have the ability to simulate quantum effects, without the extensive and complex computational logic that a classical computer would need. Therefore, quantum computers offer an avenue to explore processes that contain a large number of shower particles, thus requiring complicated parton histories and computing power, not currently achievable with modern classical techniques. Beyond QCD parton showers, this feature of a quantum computing algorithm can be of particular interest for cosmic-ray air showers, where millions of long-lived particles are simulated \cite{Wentz:2003bp,Schichtel:2019hfn}.

\subsection{Results of parton shower}\label{sec: showerresults}

\begin{figure}[ht!]
\centering
\begin{subfigure}{0.89\textwidth}
\centering
\includegraphics[width=1\textwidth]{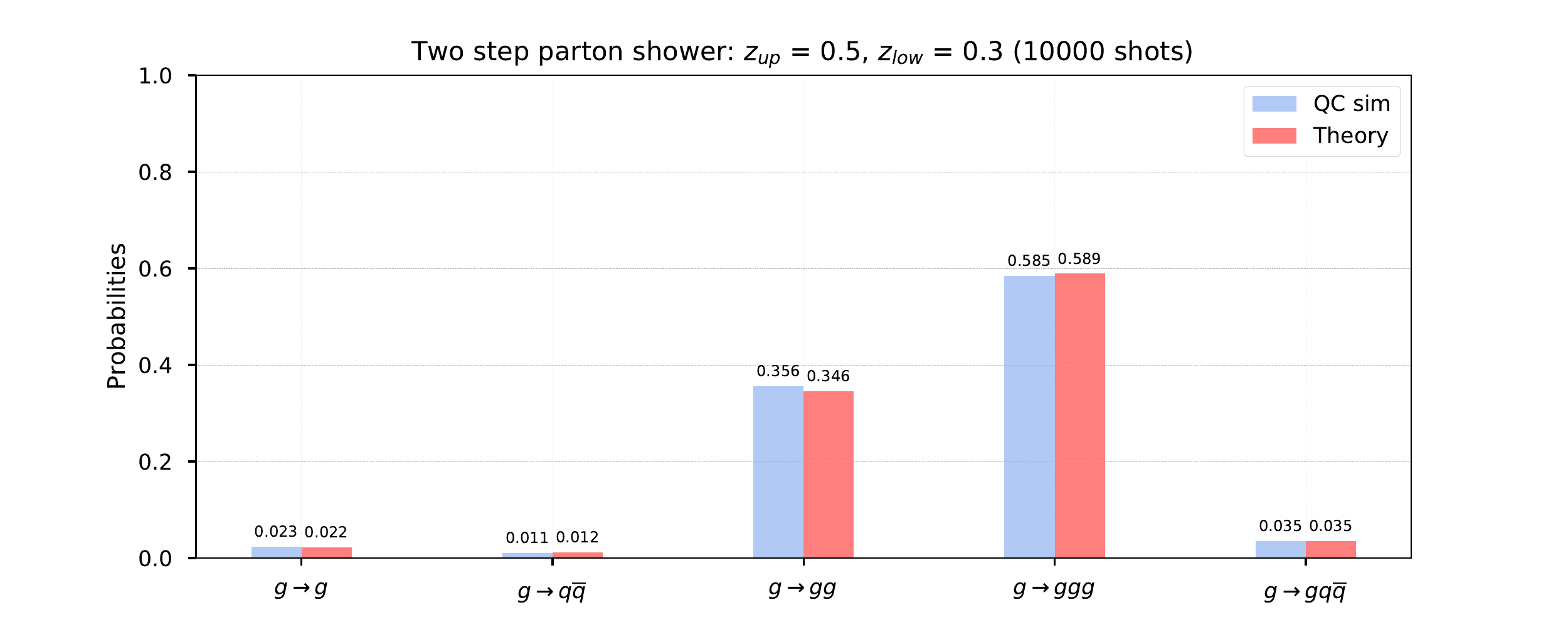}
\subcaption{Initial particle a gluon.}
\end{subfigure}
\begin{subfigure}{0.89\textwidth}
\centering
\includegraphics[width=\textwidth]{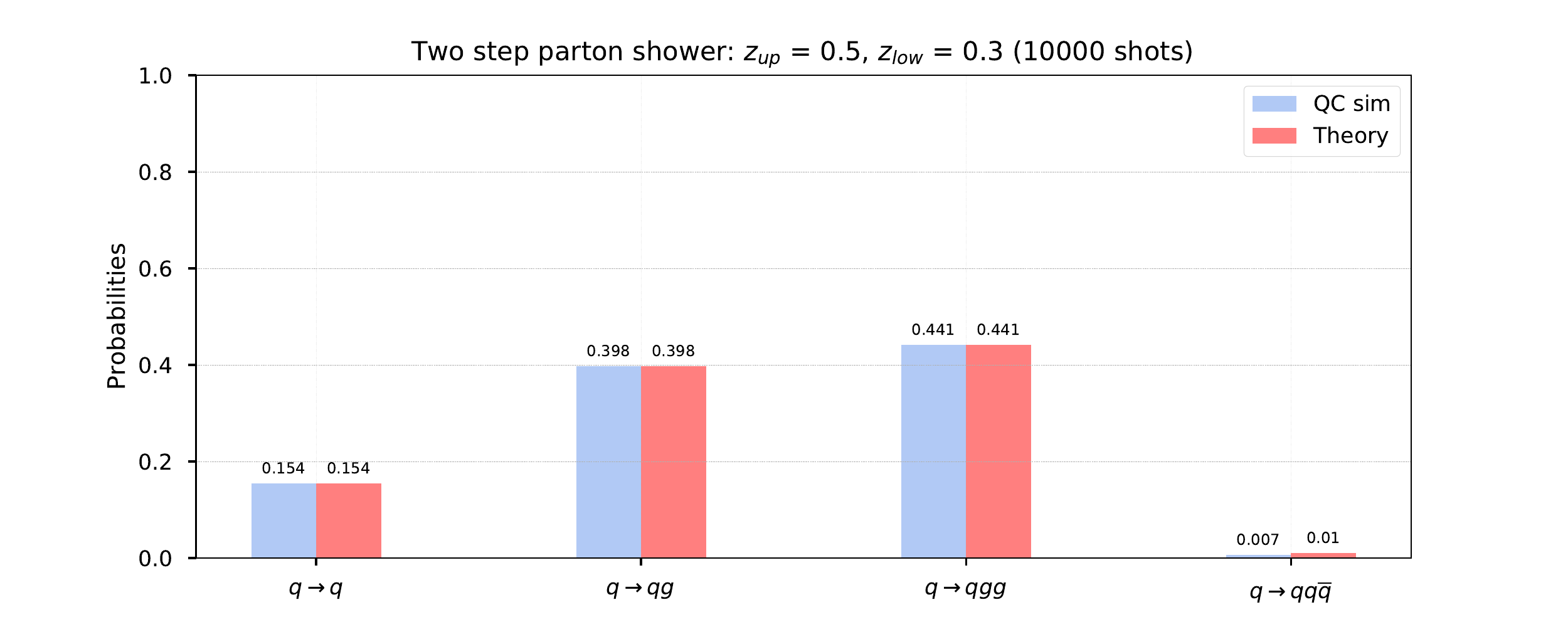}
\subcaption{Initial particle a quark.}
\end{subfigure}
\begin{subfigure}{0.89\textwidth}
\centering
\includegraphics[width=\textwidth]{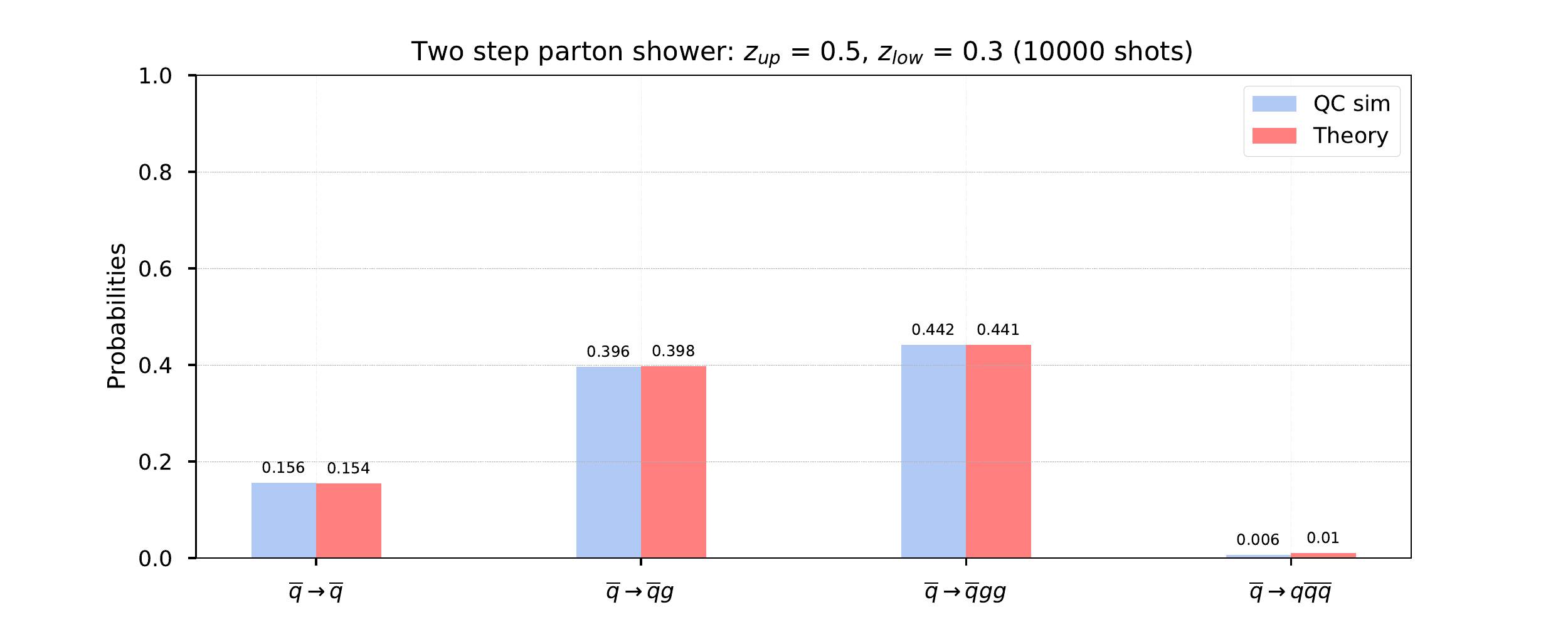}
\subcaption{Initial particle an antiquark.}
\end{subfigure}                            
\caption{Results from the quantum circuit compared to theoretical predictions for two steps of the parton shower with momentum interval of $z_{\textrm{lower}}$~=~0.3 to $z_{\textrm{upper}}$~=~0.5 and the initial state particle of (a) gluon, (b) quark and (c) antiquark. }
\label{fig:2step}
\end{figure}

A comparison of the output from the parton shower algorithm and theoretical predictions of the splitting probabilities is made, and the results are shown in Fig.~\ref{fig:2step}. The algorithm was run for 10,000 shots using the IBM Q 32-qubit Quantum Simulator~\cite{32_sim}, with a momentum interval of $z_{\textrm{lower}}$~=~0.3 to $z_{\textrm{upper}}$~=~0.5, and no noise simulation. Here the theoretical predictions have been calculated using the collinear splitting functions from Eqs.~(\ref{eqn: quarkSplittings}) and (\ref{eqn: gluonSplittings}), using the method outlined in Sec.~\ref{sec: showertheory}. The $z$ value used for the particle splitting probabilities from Eq.~(\ref{eqn: probCalc}) is the mid-point of the momentum interval used in the algorithm. The results are in agreement with the theoretically calculated probabilities to within 1$\sigma$.

A consequence of running the algorithm on a quantum simulator is that there will be no noise in the results, unlike a real quantum computer. Therefore, problems with the algorithm can be identified through direct comparison with the theoretical calculations. 
In the future, if the algorithm can be run on a real quantum computer with enough qubits, then IBM Q offers a range of noise reducing schemes for its devices through the Qiskit software \cite{IBMQ}. Another advantage of using the quantum simulator is that it automatically chooses an optimum qubit setup. In a real quantum computer, the user has to select a qubit mapping in order to optimise the operation of the computer. For future use of the algorithm, this can be done using the calibration data provided by IBM Q.

\section{Summary and conclusions}
\label{sec:conclusions}

The accurate modelling of the complexity of collisions at experiments, such as the Large Hadron Collider, relies on theoretical calculations of multi-particle processes. Such calculations can be factorised into: the hard interaction, which models the large momentum transfer, and the parton shower, which models the evolution from the hard interaction scale down to the hadronisation scale.

We present general and extendable quantum computing algorithms that provide calculations of the hard interaction process and the parton shower, as a first step towards a quantum computing algorithm to describe the full collision event at the LHC.

The hard interaction calculation uses helicity amplitudes by exploiting the equivalence of spinors and qubits, and encoding operators as a series of unitary operations in the quantum circuit, thus demonstrating an excellent use case of quantum computers to model the intrinsic quantum behaviour of the system. A quantum algorithm is constructed for two simple examples of helicity calculations; a $gq\bar{q}$ vertex and the $q\overline{q}\rightarrow q\overline{q}$ process. By applying Hadamard gates to helicity registers, we introduce a superposition state between the helicity qubits and can therefore calculate the positive and negative helicities of each particle involved simultaneously. This is further exploited in the simultaneous computation of s and t-channel amplitudes for the $q\bar{q}\rightarrow q\bar{q}$ process, thus fully utilising the quantum nature of the computation. A comparison between the theoretical predictions and the output of the quantum algorithm shows very good agreement. Furthermore, the successful implementation of the $gq\bar{q}$ vertex algorithm on a real machine is also demonstrated by comparing results from the machine with a simulator. 

We also present a quantum algorithm for simulating collinear emission in a two-step, discrete parton shower with a maximum of three final state particles, utilising the quantum computer’s ability to remain in a quantum state throughout the simulation. In contrast to classical implementations of parton showers, where individual shower histories have to be stored on a physical memory device, our quantum computing algorithm constructs a wavefunction for the whole parton shower process, which contains a superposition of all shower histories. As a result, we do not need to keep track of individual shower histories explicitly, and a physical quantity of the shower process can be obtained through a measurement of the wavefunction. The results from the algorithm, as performed on the IBM Q 32-qubit Quantum Simulator~\cite{32_sim}, show good agreement with theoretical predictions. The algorithm builds on previous work~\cite{bauer2019quantum} by including a vector boson and boson splittings, which leads to significant changes in its implementation. The ability to simulate gluon and quark splittings makes the algorithm presented here well suited to hadronic parton shower simulation and provides the foundations for developing a general parton shower algorithm. With advancements in quantum technologies, this algorithm can be extended to include all flavours of quarks without adding disproportionate computational complexity. 

With IBM recently setting their goal of exceeding 1,000 qubits by 2023~\cite{ibm_future} and advancements in the development of devices with better Quantum Volume~\cite{Jurcevic2020DemonstrationOQ}, we are on the brink of a quantum revolution. These developments would allow the algorithms presented in this paper to be extended to reflect the processes seen in experiments such as the LHC. The consequence of such advancements would be algorithms that can fully model the dynamics of quantum field theories to provide accurate and precise helicity amplitude calculations and simulations of parton showers.

\vskip 2 \baselineskip

\noindent {\it{{\bf Acknowledgements:}~~We would like to acknowledge the use of the IBM Q for this work. 
We are grateful to the authors of \cite{bauer2019quantum} for answering questions on the circuit presented in their work and for sharing their preliminary codes. M.S would like to thank Steve Abel and Daniel Maitre for helpful discussions. K.B and M.S are supported by the STFC under grant ST/P001246/1. S.M and S.W are supported by a grant from the Royal Society.}}

\vspace{1.0cm}
  
\newpage

\appendix
\section{Quantum logic gate definitions}\label{app: definitions}

\begin{itemize}
	\item \textbf{NOT gate}
		\begin{itemize}
			\item a NOT gate is a single qubit operation which flips the state of the qubit.
			
			\begin{align}
				\textrm{NOT} \vert 0 \rangle &= \vert 1 \rangle, &\textrm{NOT} \vert 1 \rangle = \vert 0 \rangle. \nonumber
			\end{align}
			The circuit representation of a NOT gate is:
			
				\begin{figure}[h!]
				\centering
				\includegraphics[scale = 0.5]{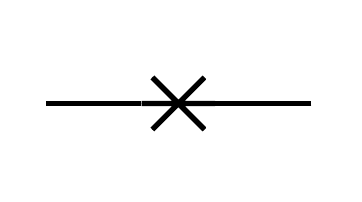}
				\end{figure}

		\end{itemize}
	\item \textbf{CNOT gate}
		\begin{itemize}
			\item a \textit{controlled}-NOT (CNOT) gate is a two qubit operation which flips the state of a target qubit dependent on the state of a control qubit.			
			\begin{align}
				\textrm{CNOT} \vert 0 0 \rangle &= \vert 0 0 \rangle, &\textrm{CNOT} \vert  0 1 \rangle = \vert 0 1 \rangle, \nonumber \\
				\textrm{CNOT} \vert 1 0 \rangle &= \vert 1 1 \rangle, &\textrm{CNOT} \vert  1 1 \rangle = \vert 1 0 \rangle. \nonumber
			\end{align}
			Here, the first qubit is the control. The circuit representation of a CNOT gate is:
			
				\begin{figure}[h!]
				\centering
				\includegraphics[scale = 0.5]{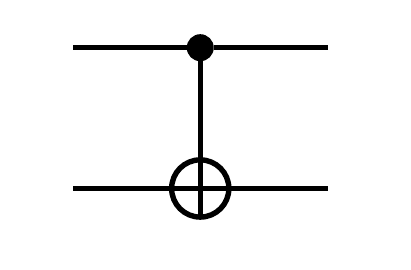}
				\end{figure}
		\end{itemize}
		
	\item \textbf{Toffoli gate (CCNOT)}
	
		\begin{itemize}
			\item A Toffoli gate is a three qubit operation, which is just a further extension of the NOT gate with two control qubits.
						
			\begin{align}
				\textrm{CCNOT} \vert 0 0 0 \rangle &= \vert 0 0 0 \rangle, &\textrm{CCNOT} \vert  0 0 1 \rangle = \vert 0 0 1 \rangle, \nonumber \\
				\textrm{CCNOT} \vert 1 0 0 \rangle &= \vert 1 0 0 \rangle, &\textrm{CCNOT} \vert  0 1 0 \rangle = \vert 0 1 0 \rangle, \nonumber \\
				\textrm{CCNOT} \vert 1 1 0 \rangle &= \vert 1 1 1 \rangle, &\textrm{CCNOT} \vert  1 1 1 \rangle = \vert 1 1 0 \rangle. \nonumber
			\end{align}
			The circuit representation of a Toffoli gate is:
			
				\begin{figure}[h!]
				\centering
				\includegraphics[scale = 0.5]{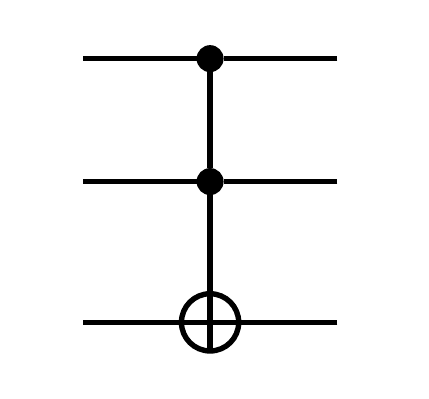}
				\end{figure}
		\end{itemize}
\newpage
	
	\item \textbf{Hadamard gate}
		\begin{itemize}
			\item a Hadamard gate is a purely quantum logic gate and does not have a classical logic gate equivalent. A Hadamard gate is a single qubit operation which puts a qubit into a superposition.
			
			\begin{align}
				\textrm{H} \vert 0 \rangle &= \frac{1}{\sqrt{2}} \big( \vert 0 \rangle + \vert 1 \rangle \big), &\textrm{H} \vert 1 \rangle &= \frac{1}{\sqrt{2}} \big( \vert 0 \rangle - \vert 1 \rangle \big). \nonumber
			\end{align}
			The Hadamard gate can be controlled, and so is only applied depending on the state of the control qubit. The circuit representation of a Hadamard gate is:
			
				\begin{figure}[h!]
				\centering
				\includegraphics[scale = 0.5]{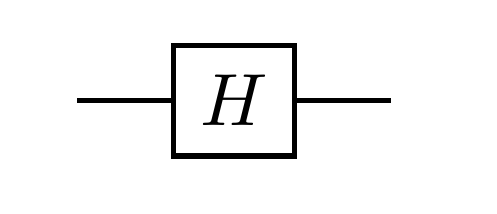}
				\end{figure}
		\end{itemize}
		
\end{itemize}

\section{Dirac and helicity spinor correspondence}\label{sec: DiracHelicity}
The following demonstration of the correspondence between Dirac spinors and Helicity spinors can be seen in Chapter 2 of~\cite{elvang_huang_2015}.

Fermion and anti-fermion spinors satisfy the Dirac equations such that,
\begin{align}\label{eqn:Dirac Eqn}
(\slashed{p}+m)u(p)&=0, & (-\slashed{p}+m)\nu (p)=0.
\end{align}
where both equations have independent solutions which can be labelled by subscripts $s = \pm $. One can move to a basis where the $\pm$ denotes spin up/down along the z-axis, by ensuring that spinors $u_{\pm }$ and $\nu _{\pm }$ are eigenstates of the z-component of the spin-matrix in the rest frame. For massless fermions, $ \pm $  denotes the helicity, the projection of the spin along the momentum of the particle. These spinors are also associated with the conventional Feynman rules for external fermions, e.g. $\nu _{\pm }(p)$ for an outgoing anti-fermion and $\bar{u} _{\pm }(p)$ for an outgoing fermion.

For the massless case, the Dirac equations reduce to
\begin{align}\label{eqn: massless Dirac Eqn}
\slashed{p}\nu  _{\pm }(p)=&0 & \bar{u} _{\pm }(p)\slashed{p}=0,
\end{align}
where $\nu _{\pm }(p)$ and $u_{\pm }(p)$ are the wavefunctions associated with outgoing anti-fermions and fermions respectively. For this case the wavefunctions are related as  $u_{\pm } =\nu _{\mp }$ and $\bar{\nu }_{\pm }=\bar{u}_{\mp }$. The two independent solutions of the Dirac equations can be written as

\begin{align}\label{eqn: spinors1}
\nu _{+}(p)&=\begin{pmatrix}
\vert p]_{a}\\ 
0
\end{pmatrix}, &\nu _{-}(p)=\begin{pmatrix}
0\\ 
\vert p \rangle^{\dot{a}}
\end{pmatrix} 
\end{align}
and

\begin{align}\label{eqn: spinors2}
\bar{u} _{-}(p)&=\begin{pmatrix}
0 & \langle p \vert _{\dot{a}}
\end{pmatrix}, &\bar{u} _{+}(p)=\begin{pmatrix}
[p \vert^{a} & 0
\end{pmatrix}
\end{align}
where the angle and square spinors are 2-component spinors that satisfy the massless Weyl equation.

\section{Helcity amplitude gate decompositions}\label{sec: helicityAmpGates}

\begin{itemize}
	\item \textbf{$U_{a\rangle}$ gate}
		\begin{itemize}
			\item The $U_{a \rangle}$ takes the form of a conventional $U_3$ rotation gate,
			
				\begin{align}
				U_{a\rangle} = U_3 (\theta, \phi, \lambda) = \begin{pmatrix} \cos \big( \frac{\theta}{2} \big) & - e^{i\lambda} \sin \big( \frac{\theta}{2} \big) \\
				e^{i\phi} \sin \big( \frac{\theta}{2} \big) & e^{i(\phi + \lambda)} \cos \big( \frac{\theta}{2} \big) \end{pmatrix}.
				\end{align}
				
				Therefore, the circuit representation is just a qiskit $U_3$ rotation, 
				
				\begin{figure}[h!]
				\centering
				\includegraphics[scale = 0.5]{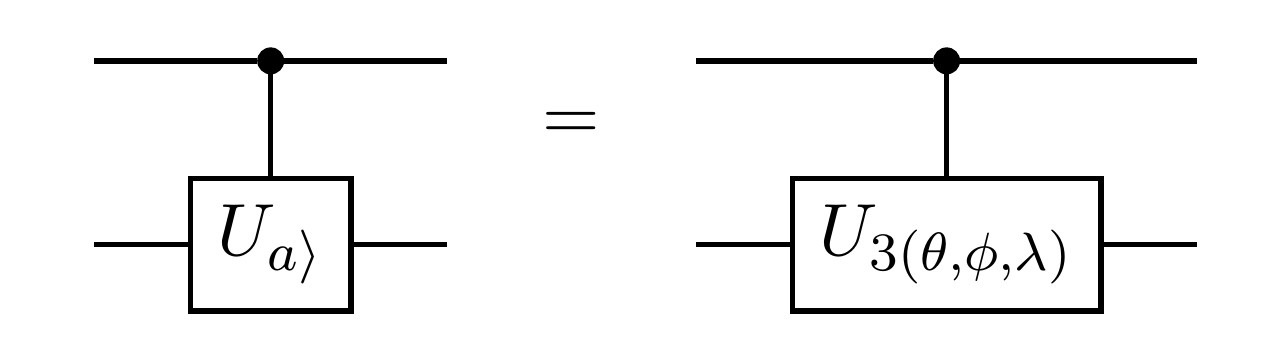}
				\end{figure}
		\end{itemize}
		
	\item \textbf{$U_{a]}$ gate}
		\begin{itemize}
			\item The $U_{a ]}$ has the matrix form,
			
				\begin{align}
				U_{a]} (\theta, \phi, \lambda) = \begin{pmatrix} - e^{-i\lambda} \sin \big( \frac{\theta}{2} \big) & e^{-i(\phi + \lambda)} 
				\cos \big( \frac{\theta}{2} \big) \\  \cos \big( \frac{\theta}{2} \big) & e^{-i\phi} \sin \big( \frac{\theta}{2} \big) \end{pmatrix}
				\end{align}
				
				Therefore, this gate has the circuit representation, 
				
				\begin{figure}[h!]
				\centering
				\includegraphics[scale = 0.5]{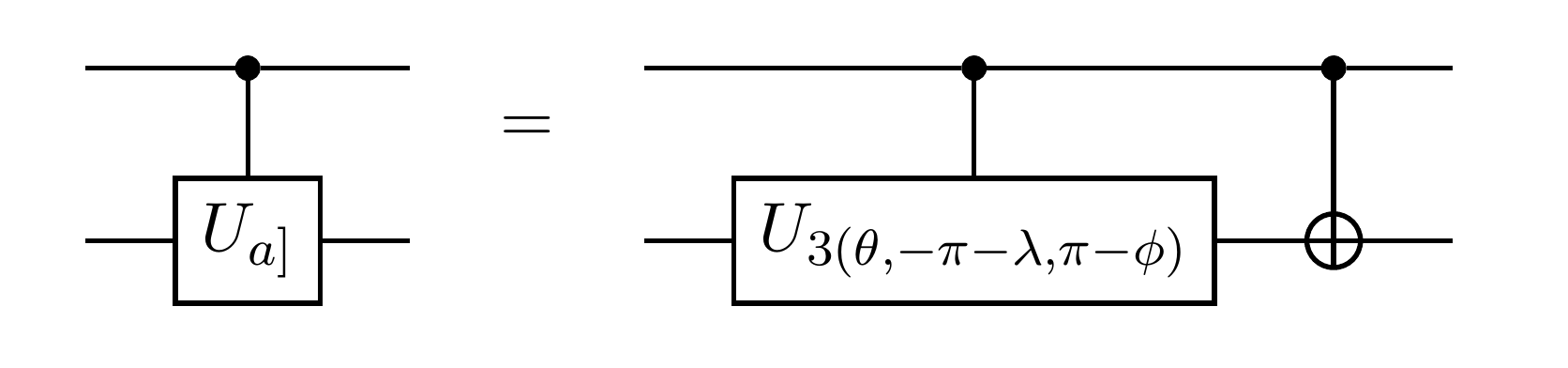}
				\end{figure}
		\end{itemize}
		
		\item \textbf{$U_{\langle b}$ gate}
		\begin{itemize}
			\item The $U_{\langle b}$ has the matrix form,
			
				\begin{align}
				U_{\langle b} (\theta, \phi) = \begin{pmatrix} - e^{i \phi} \sin \big( \frac{\theta}{2} \big) & \cos \big( \frac{\theta}{2} \big) \\
				\cos \big( \frac{\theta}{2} \big) & e^{-i \phi} \sin \big( \frac{\theta}{2} \big) \end{pmatrix}
				\end{align}
				
				Therefore, this gate has the circuit representation, 
				
				\begin{figure}[h!]
				\centering
				\includegraphics[scale = 0.5]{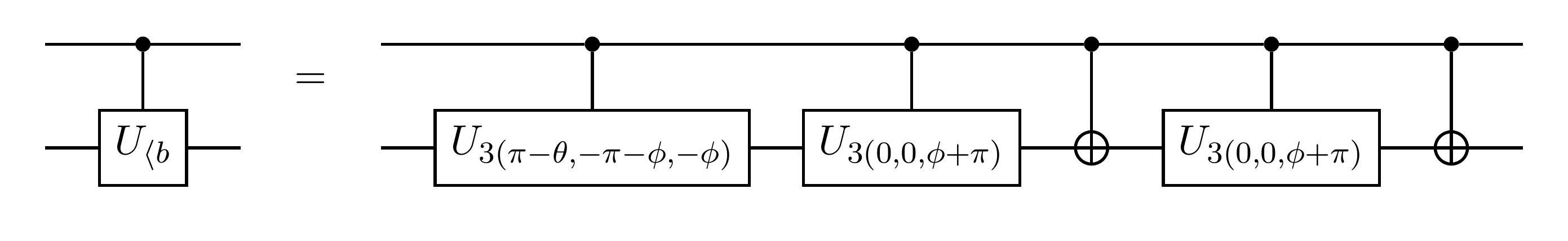}
				\end{figure}

		\end{itemize}
		\newpage
		\item \textbf{$U_{{[}b}$ gate}
		\begin{itemize}
			\item The $U_{{[}b}$ has the matrix form,
			
				\begin{align}
				U_{{[}b} (\theta, \phi) = \begin{pmatrix} \cos \big( \frac{\theta}{2} \big) & e^{-i \phi} \sin \big( \frac{\theta}{2} \big) \\
				- e^{i \phi} \sin \big( \frac{\theta}{2} \big) & \cos \big( \frac{\theta}{2} \big) \end{pmatrix}
				\end{align}
				
				Therefore, this gate has the circuit representation, 
				
				\begin{figure}[h!]
				\centering
				\includegraphics[scale = 0.5]{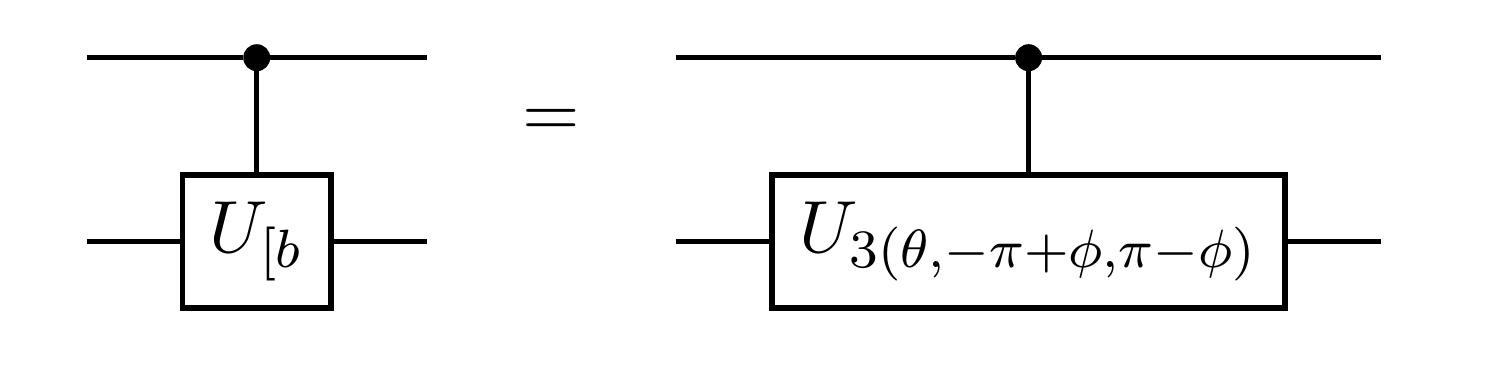}
				\end{figure}
		\end{itemize}
\end{itemize}

\section{Helicity amplitude calculation circuit diagrams and further results}\label{app:helicityAmpCalc}

\subsection{$1 \rightarrow 2$ amplitude calculation}

Here we present the detailed circuit diagram for the $q \rightarrow g\overline{q}$ process, shown in Fig.~\ref{fig: fullgqqbarCircuit}, which is implemented using the helicity amplitude gate decompositions outlined in Appendix~\ref{sec: helicityAmpGates}. This demonstrates the simplification achieved by using fully contracted helicity amplitudes in the calculation, with a scalar product calculated on each qubit. The first slice in the circuit diagram (to the left of the vertical dashed line) calculates the positive helicity, controlling from the $h$ register in the $\vert 1 \rangle$ state and the second slice (to the left of the vertical dashed line) controls from the $h$ register in the $\vert 0 \rangle$ state and calculates the negative helicity process. A superposition of both the positive and negative processes, and thus the full amplitude, is achieved by implementing a Hadamard gate on the helicity qubit, $h$.

\begin{figure}[ht!]
\centering
\includegraphics[scale = 0.5]{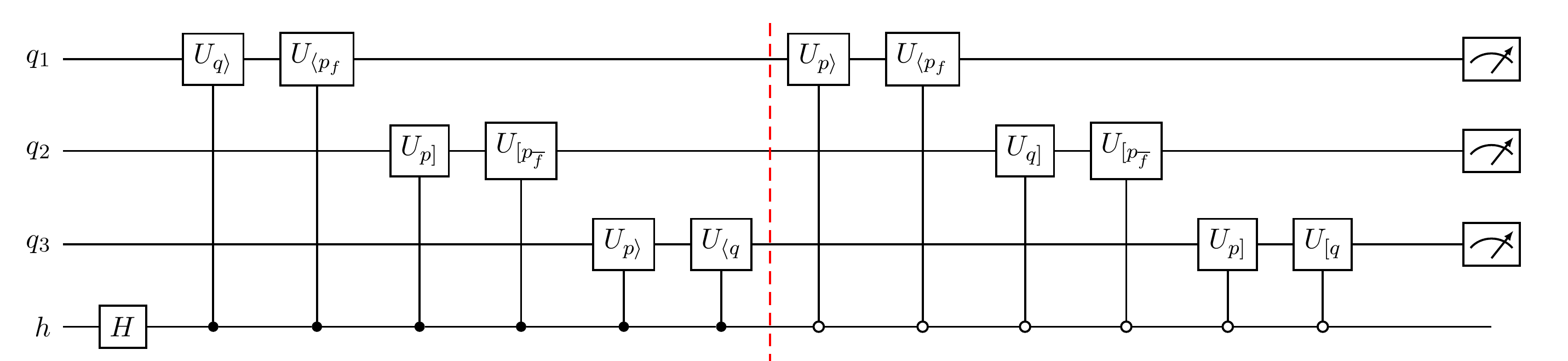}
\caption{Detailed circuit diagram for the $q \rightarrow g\overline{q}$ helicity amplitude calculation, using gate decompositions outlined in Sec.~\ref{sec: helicityAmpGates}. The amplitude for the process is calculated on the $q_i$ qubits, which are controlled from the helicity register. The $q_i$ qubits are then measured by the quantum computer.}
\label{fig: fullgqqbarCircuit} 
\end{figure}

\begin{figure}[ht!]
\centering
\begin{subfigure}{\textwidth}
\centering
\includegraphics[scale = 0.4]{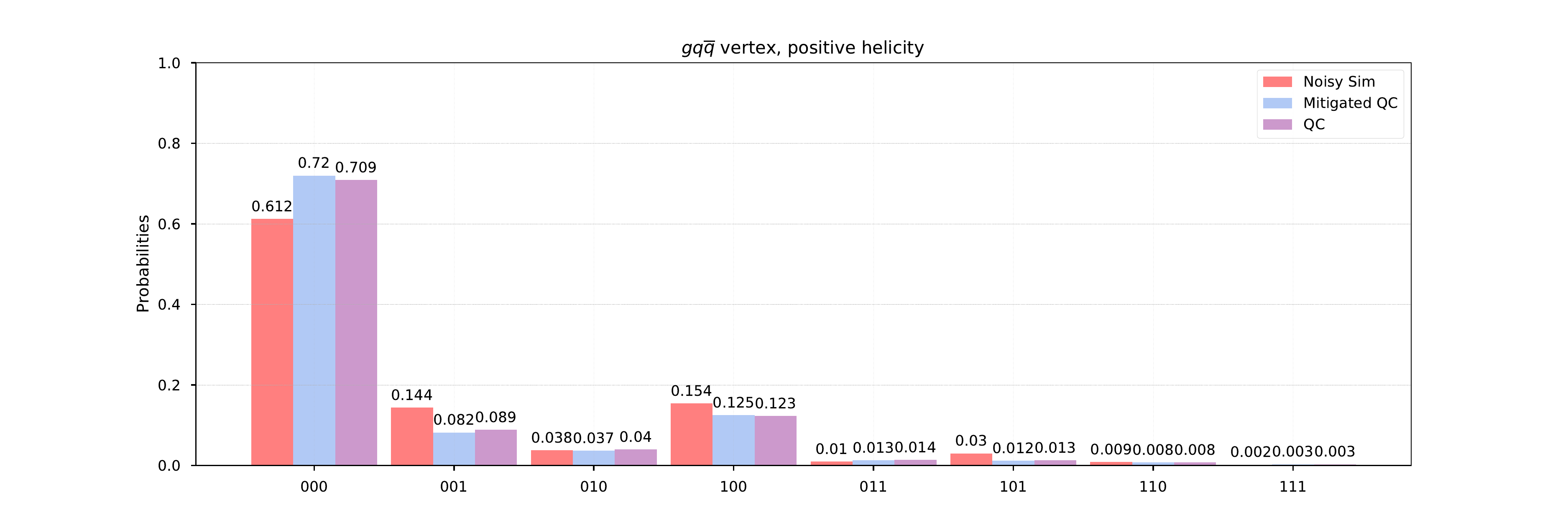}
\end{subfigure}
\begin{subfigure}{\textwidth}
\centering
\includegraphics[scale=0.4]{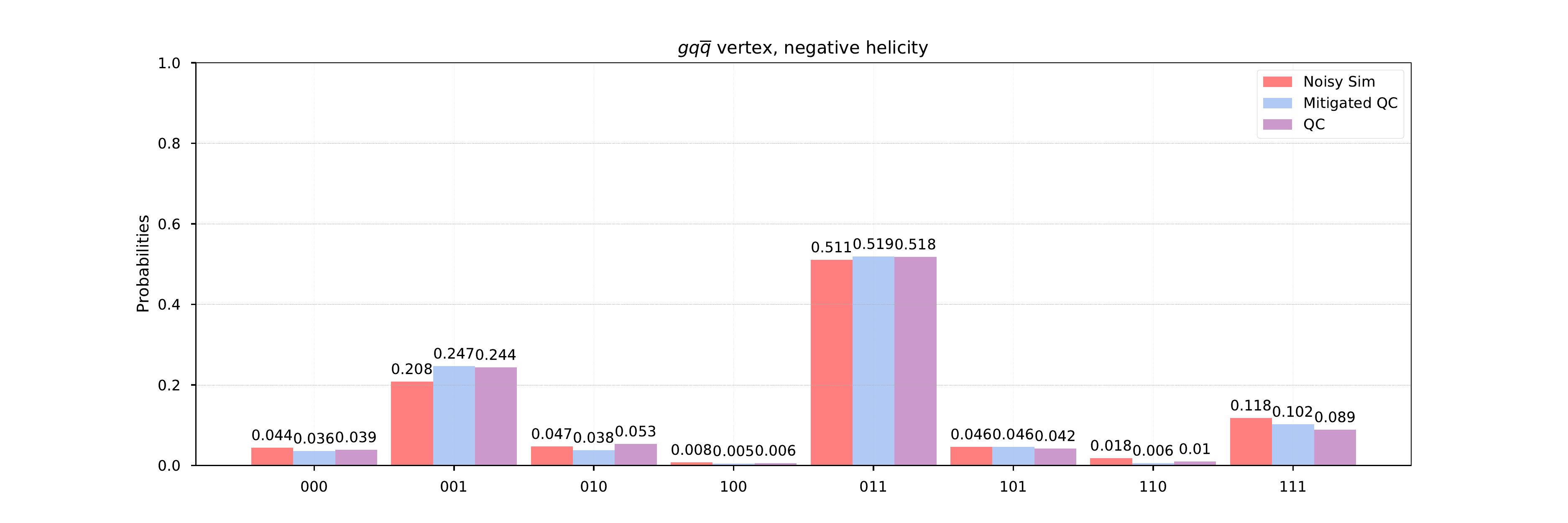}
\end{subfigure}                      
\caption{Results for the $q\rightarrow g\overline{q}$ helicity amplitude calculation. Comparison between the results from a quantum simulator with the relevant machine noise profile, the results from the Santiago quantum computer and the error mitigated results from the quantum computer. }
\label{fig:gqqbarNoise}
\end{figure}

In Fig.~\ref{fig:gqqbarNoise}, a comparison between the output of the IBM Q Santiago 5-qubit Quantum Computer~\cite{ibmq_santiago} and the IBM Q 32-qubit Quantum Simulator~\cite{32_sim} run with the Santiago device's noise profile is presented. The quantum computer has been run for 100 runs of 8192 shots, giving a total of 819,200 shots on the circuit and the simulator has been run for 10,000 shots. Here we see a more reasonable agreement between the noisy simulator and the output from the quantum computer than the comparison to the perfect machine simulator from Sec.~\ref{sec:vertex}. However, it should be noted that the noise profile used in the noisy simulation is only an approximation of the real quantum computer errors. Noise profiles are built from a limited number of parameters and are based on average measurements of qubit errors~\cite{IBMQ}. As a result, some discrepancies are present between the quantum simulator with the Santiago device's noise profile and the real device. These can be attributed to noise not accounted for in the quantum computer.

While obtaining the results, we noticed a discrepancy between separate runs on the negative helicity case. By changing the qubit mapping in the measurement process, this was identified as a tuning error on the entangling gate between qubits 2 and 3 of the Santiago machine. This error was later fixed, and the results shown were obtained from runs with a fully functioning machine. To further validate the results, a series of runs were performed on the IBM Q Valencia 5-qubit machine~\cite{ibmq_valencia}, which has a Quantum Volume of 16. The results confirmed that the Santiago machine was working correctly. 

\newpage 

\section{Detailed quantum circuit for collinear parton shower algorithm}\label{app: showercircuit}

The algorithm presented here follows a similar method to that outlined in~\cite{bauer2019quantum}. In contrast, the algorithm does not introduce flavour mixing, but does simulate a vector boson with the possibility of boson splittings. As a result, the algorithm presented here includes tailored History and Update gates to deal with the increased splitting channels. Shown in Fig.~\ref{fig: 1stepcircuit}, the circuit comprises of four tailored gate operations: Count, Emission, History, and Update gate. The particle identity is encoded using a 3-qubit base, and the following qubit combinations have been chosen for each type of particle:

\begin{equation}\label{encodedState}
\renewcommand{\arraystretch}{1.3}
\begin{tabular}{c | c c c}
\centering
& gluon & quark & antiquark \\
\hline
p $\begin{cases} p_0 \\ p_1 \\ p_2 \end{cases}$ &    $\begin{matrix} 1 \\ 0 \\ 0 \end{matrix}$   &   $\begin{matrix} 0 \\ 0 \\ 1 \end{matrix}$   &   $\begin{matrix} 0 \\ 1 \\ 1 \end{matrix}$       
\end{tabular}
\end{equation}
Using a 3-qubit base, it is possible to simulate 7 different types of particle and 1 null state. Therefore, the algorithm could be easily extended to accommodate more quark flavours if more qubits were available.

\subsection{Count gate}

\begin{figure}[ht!]
\centering
\includegraphics[scale = 0.5]{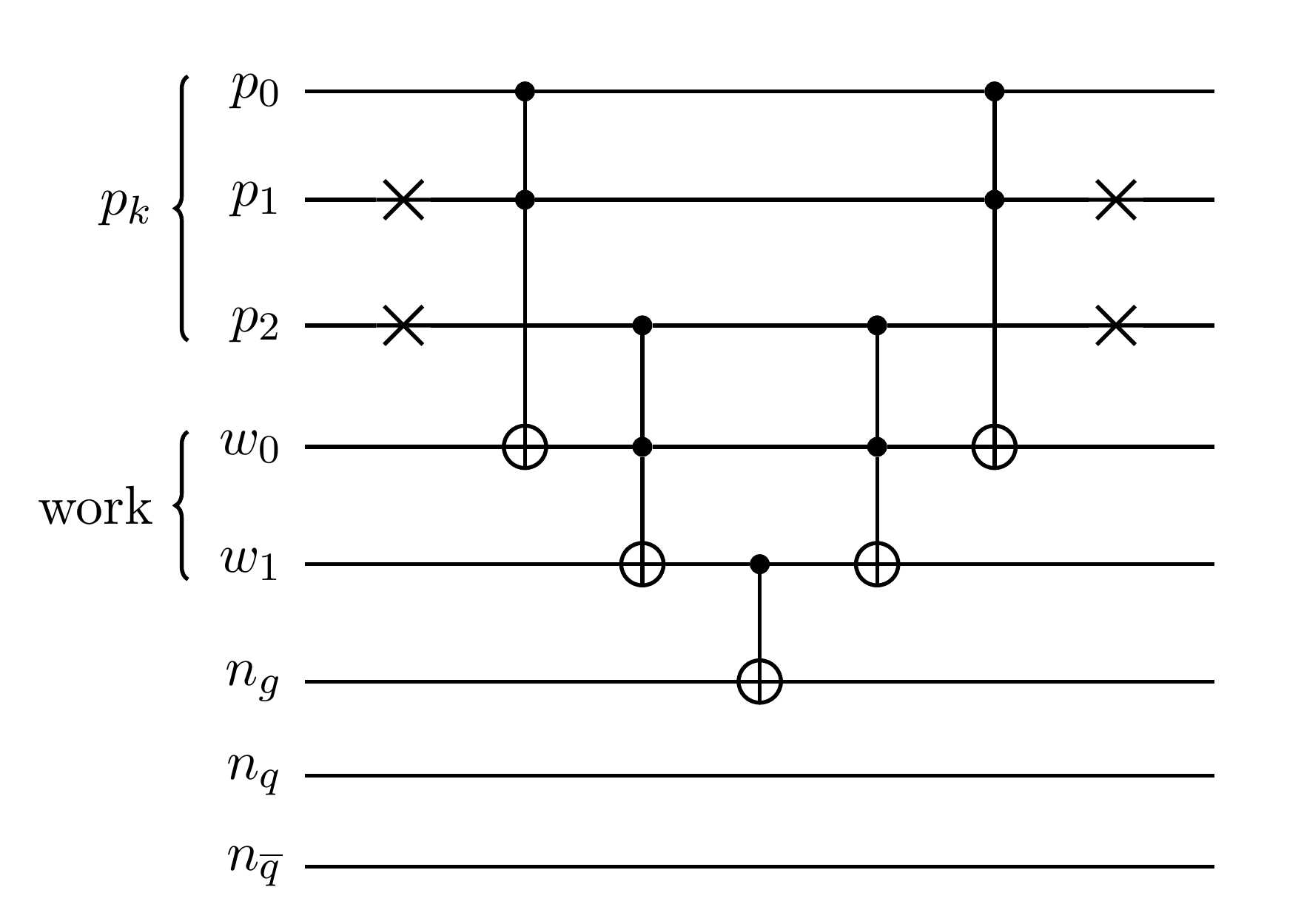}
\caption{Count gate circuit decomposition for counting a gluon in the particle register. To complete the count gate, this is repeated for all other possible particle types by applying different combinations of \textit{NOT} gates.}
\label{fig: countGate}
\end{figure}

The count gate comprises of three individual counting mechanisms for each type of particle, and is applied to each particle register individually. The algorithm utilises a series of \textit{NOT}, \textit{controlled-NOT} (\textit{CNOT}) and Toffoli (\textit{CCNOT}) gates to update the count registers, $n_i$, depending on the type of particle represented in the particle register. Fig.~\ref{fig: countGate} shows the counting mechanism for a gluon, controlling only from the gluon state outlined in \ref{encodedState}. 

The total number of count registers, $n_i$, used in the algorithm is 4. As the particle count registers are updated at the beginning of a step, the maximum number of gluons that can be present is 2, and the maximum number of quarks/antiquarks is 1. Therefore, for this algorithm, only 2 gluon count registers and 1 quark/antiquark count register are required. Ideally, one would have the same number of count registers for each particle type, which would be equal to the step number. However, due to the limited number of available qubits, this has not been possible here.

\subsection{Emission gate}

The emission gate implements the Sudakov factors from Eq.~(\ref{eqn: sudakovs}) by defining a $U_3$ rotation that can be applied to the emission register, $e$. The structure of this rotation takes the same form as that presented in~\cite{bauer2019quantum},

\begin{equation}\label{eqn: emissionRot}
U_e = \begin{pmatrix} \sqrt{\Delta_{\textrm{tot}} (z_1, z_2)} & -\sqrt{1 - \Delta_{\textrm{tot}} (z_1, z_2)} \\ \sqrt{1 - \Delta_{\textrm{tot}} (z_1, z_2)} & \sqrt{\Delta_{\textrm{tot}} (z_1, z_2)} \end{pmatrix}.
\end{equation} 
This rotation changes the state of the emission gate, $e$, to $\vert 1 \rangle$ if there is an emission, and keeps it in state $\vert 0 \rangle$ if there is no emission. Non-emission probabilities (Sudakov factors) are used due to the Qiskit \cite{IBMQ} definition of a qubit state,

\begin{align}\label{eqn: qubits}
\vert 0 \rangle &= \begin{pmatrix} 1 \\ 0 \end{pmatrix}, & \vert 1 \rangle = \begin{pmatrix} 0 \\ 1 \end{pmatrix}.
\end{align}

\begin{figure}[ht!]
\centering
\includegraphics[scale = 0.5]{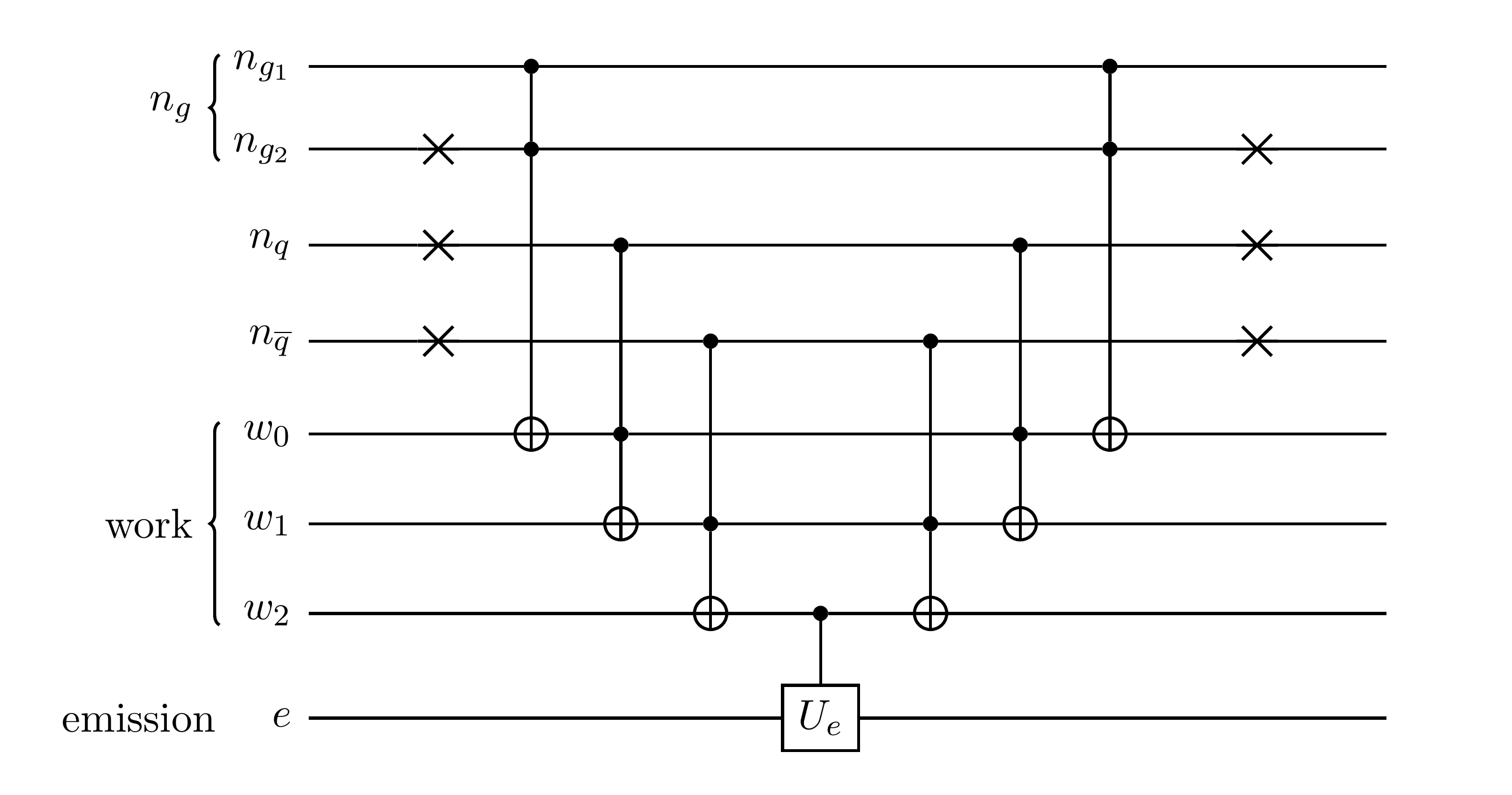}
\caption{Emission gate for a single gluon in the first particle register. Here the $U_e$ is a $U_3$ rotation is used to implement the Sudakov factors.}
\label{Fig. EmissionGate}
\end{figure}

Similarly to the Count gate, the Emission gate is constructed from a series of \textit{NOT} gates which determine the target state, and a series of \textit{CCNOT} gates which implement the operation if the target state is present. Here, the emission is determined by controlling from the particle count gates. If the desired particles are present, then the emission rotation from Eq.~(\ref{eqn: emissionRot}) is applied to the emission register. As only one emission can occur in a single step, then only one emission qubit is needed per step.

\subsection{History gate}

The history gate is the most complicated implementation in the algorithm. This is largely due to the fact that a gluon can split to either a gluon pair, or a quark-antiquark pair. As a consequence this requires two calculations of splitting probabilities for a gluon, as outlined in Eq.~(\ref{eqn: probCalc}). These probabilities are implemented by controlling from present particles and applying a rotation to the relevant history register; again taking a form similar to the one presented in~\cite{bauer2019quantum},

\begin{equation}\label{eqn: historyRot}
U_h = \begin{pmatrix} \sqrt{1 - \frac{P_{k\rightarrow ij}(z) }{P_\textrm{tot}(z)}} & -\sqrt{ \frac{P_{k\rightarrow ij}(z) }{P_\textrm{tot}(z)}} \\ \sqrt{ \frac{P_{k\rightarrow ij}(z) }{P_\textrm{tot}(z)}} & \sqrt{1 -  \frac{P_{k\rightarrow ij}(z) }{P_\textrm{tot}(z)}} \end{pmatrix},
\end{equation}
where $P_\textrm{tot}$ is defined as,
\begin{equation}
P_\textrm{tot} (z) = n_g ( P_{g \rightarrow q\overline{q}} + P_{g \rightarrow gg} ) + n_q P_{q \rightarrow qg} + n_{\overline{q}} P_{\overline{q} \rightarrow \overline{q} g}. 
\end{equation}
Here the non-splitting probabilities are used in the diagonal elements due to the definition of the qubit states outlined in Eq.~(\ref{eqn: qubits}).

\begin{figure}[ht!]
\centering
\includegraphics[scale = 0.5]{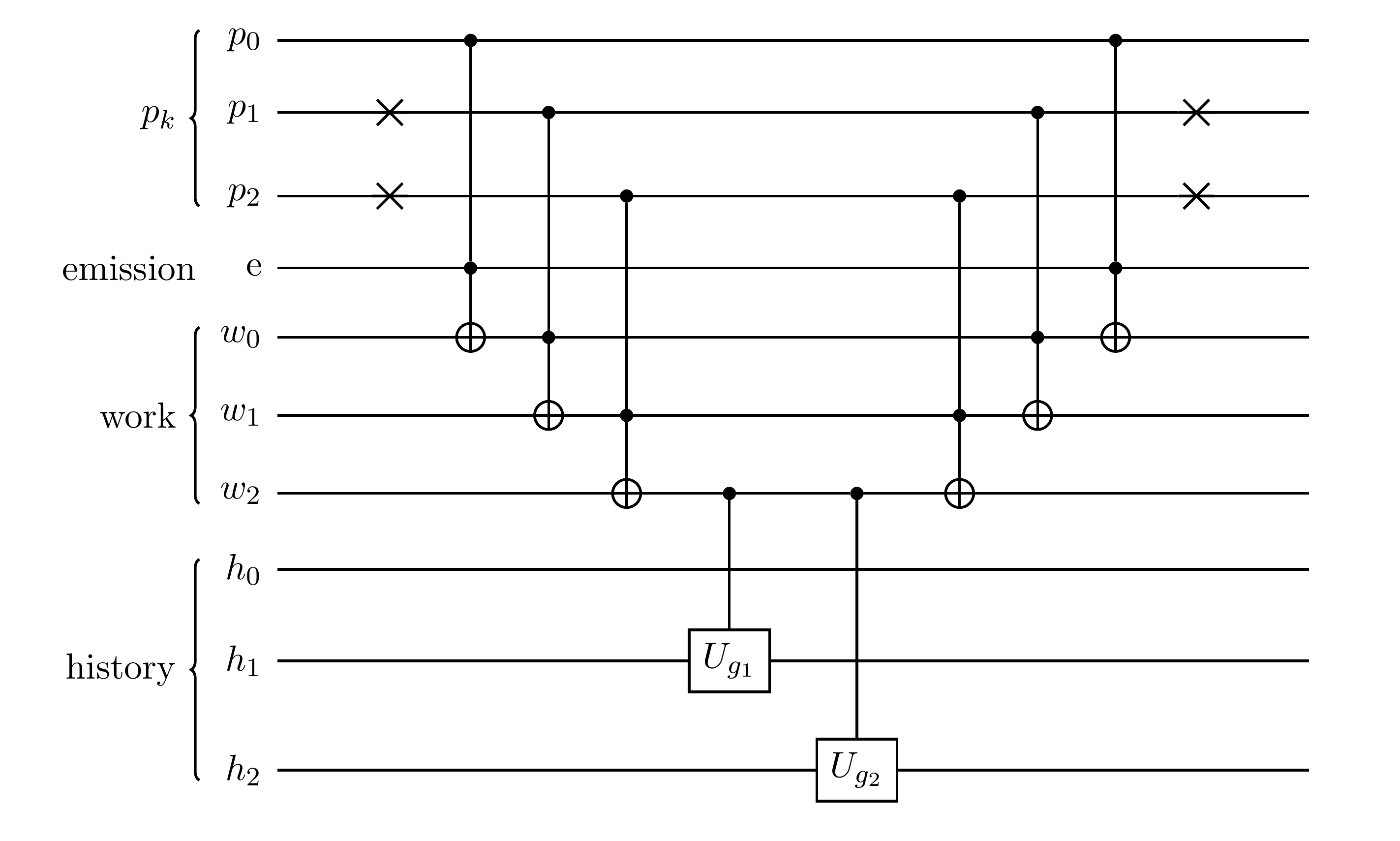}
\caption{History gate for a single gluon in the first step. Here the $U_h$ gate is a $U_3$ rotation used to implement the splitting probabilities.}
\label{fig: historyGate}
\end{figure}

The history gate used in this algorithm differs from~\cite{bauer2019quantum}, such that it controls from the particle registers and not the count registers. This is to reduce the number of count registers needed in the algorithm. For this algorithm, the history rotation needs to know which particle is being considered and which particle register it is in so that the correct rotation can be applied to the correct history qubit. This could be done by increasing the count registers by one qubit per particle type every step, to have a count register for each possible particle in a specific particle register. However, this need for more counting qubits can be reduced by simply controlling from the particle registers themselves; this is shown for a gluon in Fig.~\ref{fig: historyGate}. This can be done without impacting the rest of the algorithm, as long as there are enough count qubits to count the number of present particles correctly. This is because the emission gate does not need to know what specific particle is present in which particle register, just how many particles are present.

Once the particle content of the simulation has been assessed, the history rotations, $U_h$, from Eq.~(\ref{eqn: historyRot}) are applied to the relevant history registers. The first, labelled $g_1$, is for the $g\rightarrow q\overline{q}$ splitting, and the second, labelled $g_2$, is for the $g\rightarrow gg$. Note that both of these rotations could result in a splitting, and thus rotate the history qubit to the $\vert 1 \rangle$ state. Therefore, they are applied to different history registers. In general, one could have a condition on the second rotation, such that it is not applied if the first rotation yields a $\vert 1\rangle$, but in this algorithm, this condition was carried forward to the update gate, see Sec.~\ref{app: updateGate}. As a result of these different splittings, the required number of qubits needed for the history register in each step is $3N$, where $N$ is the step number. Figure~\ref{fig: historyGate} shows the history gate for the first step, thus 3 qubits are needed for the history register: two for the gluon splittings, and one for the quark/antiquark splittings.

\subsection{Update gate}\label{app: updateGate}

\begin{figure}[ht!]
\centering
\includegraphics[scale = 0.5]{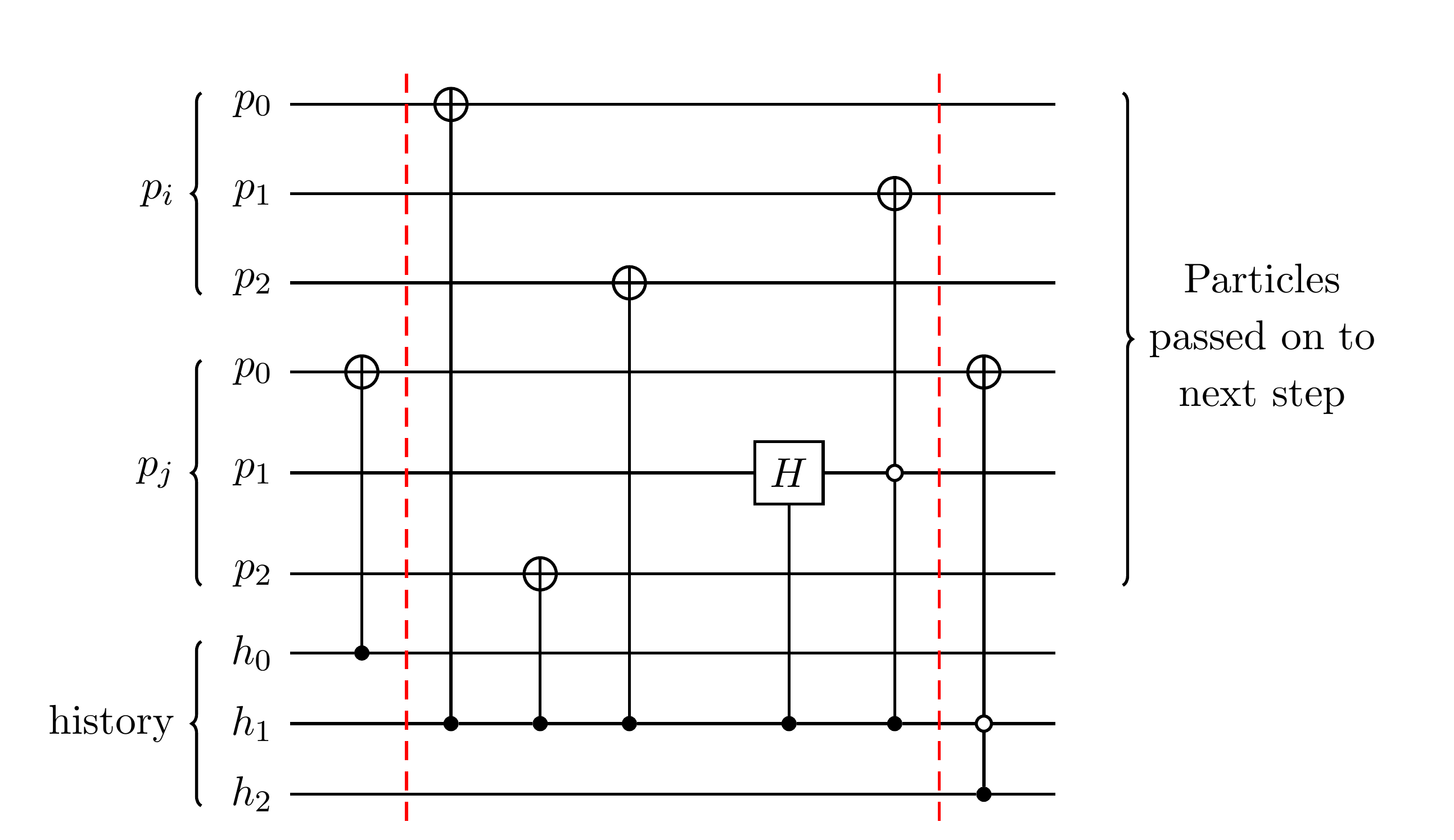}
\caption{Update gate for the first step of the algorithm. Each slice is a different update mechanism: far left slice updates $q\rightarrow qg$ splittings, centre slice updates $g \rightarrow q\overline{q}$ and the far right slice updates $g\rightarrow gg$.}
\label{fig: updateGate}
\end{figure}

The final gate in the algorithm is the update gate, which, if an emission has occurred, changes the particle content of the particle registers accordingly. Figure~\ref{fig: updateGate} shows the update gate from the first step, which is the simplest update gate in the algorithm. The circuit shown is sliced into individual updates. The first slice on the left shows the addition of a new gluon to the particle register. This is controlled from the quark/antiquark history gate, and so corresponds to the $\overset{\scriptscriptstyle(-)}{q}\rightarrow\overset{\scriptscriptstyle(-)}{q}g$ process. 

The middle slice in Fig.~\ref{fig: updateGate} shows the update of a gluon splitting to a quark-antiquark pair controlled from the $g\rightarrow q\overline{q}$ history register. The first three \textit{CNOT} gates of this slice put the particle registers into a state of two quarks. The update gate then utilises a \textit{controlled-Hadamard} gate, putting the $p_{j_1}$ qubit in a superposition of $\vert 0 \rangle$ and $\vert 1\rangle$ states. The final gate of the slice controls from the history register, but also controls on a $\vert 0 \rangle$ state on the $p_{j_1}$. At the point of measurement, if $p_{j_1}$ is measured as a $\vert 0 \rangle$ state, then the $p_k$ register represents an antiquark and $p_j$ represents a quark. If $p_{j_1}$ is measured in the $\vert 1 \rangle$ state, then the $p_k$ register represents a quark and the $p_j$ register represents an antiquark. In a sense, this \textit{controlled-Hadamard} gate and subsequent \textit{CCNOT} gate put the particle registers $p_j$ and $p_k$ into a superposition of quark-antiquark and antiquark-quark states.

The final slice on the right of the circuit diagram in Fig.~\ref{fig: updateGate} shows the update gate corresponding to the $g\rightarrow gg$ process. This simple update changes the $p_{j_0}$ qubit to a $\vert 1 \rangle$ state controlled from the history register, like the quark/antiquark update gate. However, this is where the algorithm adds a preference to $g\rightarrow q\overline{q}$ process over the $g\rightarrow gg$ process. The \textit{CCNOT} gate for the final slice in Fig.~\ref{fig: updateGate} also controls from a $\vert 0 \rangle$ state on the $g\rightarrow q\overline{q}$ history qubit. Therefore a gluon can only split to a gluon pair if the history gate for a gluon splitting to a quark-antiquark pair is in the $\vert 0 \rangle$ state. This is an acceptable approximation because the splitting probabilities for $g \rightarrow q\overline{q}$ are a lot less than for $g \rightarrow gg$. Consequently, there is only a small probability that they are both in the $\vert 1 \rangle$ state at any one time. However, it is possible that this may be a limitation in comparison to current classical parton shower algorithms provided by packages such as \textsc{Pythia}~\cite{Pythia}, \textsc{Herwig}~\cite{Herwig}, and \textsc{Sherpa}~\cite{Gleisberg:2003xi}, as these give more complex weightings to the different splitting channels.

  
\bibliographystyle{JHEP}
\bibliography{references}

\end{document}